\documentclass[11pt]{article}

\usepackage{fullpage}
\usepackage{setspace}
\usepackage{amsmath}
\usepackage{amsfonts}
\usepackage{amssymb}
\usepackage{amsthm}
\usepackage{mathtools}

\usepackage{authblk}

\usepackage{graphicx}
\usepackage{algorithm}
\usepackage{algpseudocode}

\usepackage[round]{natbib}
\usepackage{xcolor}
\usepackage[
    colorlinks=true,
    linkcolor=blue,
    urlcolor=blue,
    citecolor=blue
]{hyperref}

\newtheorem{lemma}{Lemma}
\newtheorem{remark}{Remark}

\newcommand{\indep}{\perp\!\!\!\perp}

\title{Empirical Bayes Method for Large-Scale Multiple Testing with Heteroscedastic Errors}

\author[1]{Kwangok Seo}
\author[1]{Johan Lim\thanks{Corresponding author. Email: johanlim@snu.ac.kr}}
\author[2]{Kaiwen Wang}
\author[3]{Dohwan Park}
\author[4]{Shota Katayama}
\author[5]{Xinlei Wang}

\affil[1]{Department of Statistics, Seoul National University, Seoul, Korea}
\affil[2]{Department of Data Science, Davidson College, NC, USA}
\affil[3]{Department of Mathematics and Statistics, University of Maryland at Baltimore County, MD, USA}
\affil[4]{Faculty of Economics, Keio University, Tokyo, Japan}
\affil[5]{Department of Mathematics, University of Texas at Arlington, TX, USA}

\date{}

\begin{document}

\maketitle 

\begin{abstract}
    In this paper, we address the normal mean inference problem, which involves testing multiple means of normal random variables with heteroscedastic variances. Most existing empirical Bayes methods for this setting are developed under restrictive assumptions, such as the scaled inverse-chi-squared prior for variances and unimodality for the non-null mean distribution. However, when either of these assumptions is violated, these methods often fail to control the false discovery rate (FDR) at the target level or suffer from a substantial loss of power. To overcome these limitations, we propose a new empirical Bayes method, gg-Mix, which assumes only independence between the normal means and variances, without imposing any structural restrictions on their distributions. We thoroughly evaluate the FDR control and power of gg-Mix through extensive numerical studies and demonstrate its superior performance compared to existing methods. Finally, we apply gg-Mix to three real data examples to further illustrate the practical advantages of our approach.
    
\end{abstract}

\section{Introduction}

\subsection{Normal Mean Inference Problem}
Consider a sequence of pairs of random variables $\{(X_i,S_i^2)\}_{i=1}^m$, with each pair distributed as
\begin{equation}\label{eqn:like}
    (X_i,S_i^2)
    \sim
    N(\mu_i,\sigma_i^2)
    \otimes
    \frac{\sigma_i^2}{\nu}\chi_\nu^2,
    \qquad i\in[m] \coloneqq\{1,\ldots,m\},
\end{equation}
where $N(\mu, \sigma^2)$ denotes the normal distribution with mean $\mu$ and variance $\sigma^2$, $\chi^2_\nu$ denotes the chi-squared distribution with $\nu$ degrees of freedom, and $\otimes$ denotes the product of measures, indicating that the two random variables are assumed to be independent. The quantity $X_i$ is an estimator of the parameter of interest $\mu_i$, while $S_i^2$ is an estimator of its variance $\sigma_i^2$.

This paper addresses the normal mean inference problem of simultaneously testing the following $m$ hypotheses:
\begin{equation}\label{eqn:hypo}
    H_{0,i} : \mu_i = 0
    \quad \text{vs.} \quad
    H_{1,i} : \mu_i \neq 0,
    \qquad i \in [m].
\end{equation}
Each hypothesis is associated with a summary statistic pair $(X_i, S_i^2)$, and our objective is to test the hypotheses in \eqref{eqn:hypo} using these statistics.

This type of problem commonly arises in the analysis of gene expression data, where the objective is to identify genes that are differentially expressed between two groups: the treatment and control groups.
Specifically, consider an $ m \times n $ data matrix $ \mathcal{D} $, where $ m $ represents the number of genes and $ n $ is the total number of subjects. The $(i, j)$-th element in $ \mathcal{D} $ corresponds to the (preprocessed) expression level of the $ i $-th gene in the $ j $-th subject. Without loss of generality, we assume that the first $ n_1 $ subjects belong to the treatment group and the remaining $ n_2 = n - n_1 $ subjects belong to the control group. Then, the $ i $-th row of the data matrix can be written as
\begin{equation*}
    \boldsymbol{d}_i^\top = (a_{i1}, a_{i2}, \dots, a_{in_1}, b_{i1}, b_{i2}, \dots, b_{in_2}).
\end{equation*}
We model the gene expression levels as
\begin{align}
    \begin{split}\label{eqn:dat_gen}
        a_{ij} = \mu_{ai} + \epsilon_{aij}, \quad \epsilon_{aij} \overset{\mathrm{i.i.d.}}{\sim} N(0, \eta_i^2), \quad j \in [n_1],\\
        b_{ij} = \mu_{bi} + \epsilon_{bij},\; \quad \epsilon_{bij} \overset{\mathrm{i.i.d.}}{\sim} N(0, \eta_i^2), \quad j \in [n_2].
    \end{split}
\end{align}
The primary object of interest is the difference in mean expression levels between the treatment and control groups, defined as $\mu_i = \mu_{ai} - \mu_{bi}$. A natural estimator of $\mu_i$ is the difference in sample means, 
\begin{equation*}
    X_i = \bar{a}_i - \bar{b}_i,
\end{equation*} 
where $\bar{a}_i$ and $\bar{b}_i$ denote the sample means of the treatment and control groups, respectively, for the $i$-th gene.
A natural estimator of the variance of $X_i$ is given by
\begin{equation*}
    S_i^2 = \frac{(n_1 - 1) S_{ai}^2 + (n_2 - 1) S_{bi}^2}{n_1 + n_2 - 2} \left( \frac{1}{n_1} + \frac{1}{n_2} \right),
\end{equation*}
where $S_{ai}^2$ and $S_{bi}^2$ denote the sample variances of the treatment and control groups, respectively, for the $i$-th gene.
Under the data generation model described in \eqref{eqn:dat_gen}, $X_i$ follows a normal distribution with mean $\mu_i=\mu_{ai}-\mu_{bi}$ and variance $\sigma_i^2=\eta_i^2\left(\frac{1}{n_1}+\frac{1}{n_2}\right)$, while $S_i^2$ follows a scaled chi-squared distribution with degrees of freedom $\nu=n_1+n_2-2$ and scale parameter $\sigma_i^2/\nu$. Moreover, under the Gaussian sampling model, the sample means are independent of the corresponding sample variances, which implies that $X_i$ and $S_i^2$ are independent. Therefore, the pair $(X_i,S_i^2)$ satisfies the distributional assumption in \eqref{eqn:like}.
Therefore, the problem of identifying differentially expressed genes between the two groups reduces to the problem of testing the hypotheses in \eqref{eqn:hypo}. 

\subsection{Existing Methods for Normal Mean Inference Problem}
Various multiple testing procedures have been proposed for the normal mean inference problem. 
In this paper, we restrict our attention to methods developed under the assumption that the pairs $\{(X_i,S_i^2)\}_{i=1}^m$ are mutually independent.
In what follows, we review representative methods classified according to the modeling assumptions for $\mu_i$ and $\sigma_i^2$.

One line of work builds on the empirical partially Bayes framework \citep{cox1975note}. In this framework, the nuisance parameter $\sigma_i^2$ is modeled as a random variable, while the parameter of interest $\mu_i$ is regarded as a fixed but unknown constant. Methods in this class are distinguished primarily by the prior assumption on $\sigma_i^2$. For example, \cite{smyth2004linear} assumes that the prior distribution of $\sigma_i^2$ belongs to the scaled inverse-chi-squared family. Under this assumption, the conditional null distribution of $X$ given $S^2$ is derived. Based on this distribution, a conditional $p$-value is computed for each hypothesis. These conditional $p$-values are then input into the Benjamini--Hochberg (BH) procedure \citep{benjamini1995controlling} to determine the rejection set. However, this approach relies on a parametric assumption, which can be overly restrictive in some applications. To address this limitation, \cite{ignatiadis2025empirical} generalize the framework by replacing the parametric prior with a nonparametric one, allowing $\sigma_i^2$ to follow an arbitrary distribution supported on $[0,\infty)$.

Another line of work is based on the empirical Bayes framework \citep{robbins1956empirical}, in which both $\mu_i$ and $\sigma_i^2$ are treated as random variables. Methods in this class are characterized by their joint prior assumptions on $(\mu_i, \sigma_i^2)$.
For example, \cite{lu2019empirical} assume the following priors:
\begin{equation*}
	\sigma_i^2 \overset{\mathrm{i.i.d.}}{\sim} \text{Scaled-Inv-}\chi^2, \quad
	\mu_i \overset{\mathrm{i.i.d.}}{\sim} (1 - \pi)~\delta_0(\cdot) + \pi~f(\cdot),
\end{equation*}
where $\delta_0$ is the Dirac delta function at zero, $\pi$ denotes the proportion of non-null hypotheses, and $f$ is a unimodal density centered at zero. Under this specification, the local false discovery rate (Lfdr; \citealp{efron2001empirical})  is used as the ranking statistic, and hypotheses with Lfdr values below a data-adaptive threshold are rejected. To increase flexibility, \cite{zheng2021mixtwice} further generalize the prior on $\sigma_i^2$ to allow for any distribution supported on $[0, \infty)$, while retaining the unimodality assumption on $f$.

Apart from the aforementioned approaches, some methods operate under the assumption that the variance $\sigma_i^2$ is known, and therefore directly use $\sigma_i^2$ in the inference procedure instead of the sample variance $S_i^2$. For instance, both \cite{stephens2017false} and \cite{fu2022heteroscedasticity} develop methods under this setting, assuming the following prior model:
\begin{equation*}
    \sigma_i^2 \overset{\mathrm{i.i.d.}}{\sim} G(\cdot), \quad \mu_i \overset{\mathrm{i.i.d.}}{\sim} (1 - \pi)~\delta_0(\cdot) + \pi~f(\cdot),
\end{equation*}
where $G$ denotes an arbitrary distribution with support on $[0, \infty)$. The main distinction between these methods lies in their assumptions about the non-null density $f$: \cite{stephens2017false} imposes a unimodality constraint on $f$, whereas \cite{fu2022heteroscedasticity} allow $f$ to be completely unrestricted. In both cases, the Lfdr serves as the ranking statistic, and hypotheses with Lfdr
values below a data-adaptive threshold are rejected.

\subsection{Limitations of Existing Methods}
While the aforementioned methods offer useful solutions, each has certain limitations. In this section, we review these limitations in detail.

First, the empirical partially Bayes methods \citep{smyth2004linear, ignatiadis2025empirical} do not incorporate the structure of the non-null distribution of $\mu_i$, which can lead to a loss of power compared to methods that explicitly incorporate such structural information into the testing procedure (see Figure~\ref{fig:7}).

Second, the scaled inverse-chi-squared assumption for $\sigma_i^2$ \citep{smyth2004linear, lu2019empirical} can be overly restrictive, as pointed out in several studies \citep{lu2016variance, phipson2016robust}. Our numerical study also shows that procedures relying on this assumption may fail to control the type I error or suffer a loss of power when the assumption is violated (see Figures~\ref{fig:2}, \ref{fig:3}, \ref{fig:5}, and \ref{fig:6}).

Third, existing empirical Bayes methods commonly rely on the assumption that the non-null prior distribution of $\mu_i$ is unimodal \citep{stephens2017false, lu2019empirical, zheng2021mixtwice}, although this assumption cannot be verified from the observed data. We numerically show that when the true non-null distribution of $\mu_i$ is bimodal, the methods based on the unimodality assumption tend to underestimate the null proportion (see Figures~\ref{fig:1} and~\ref{fig:4}) and fail to control the type I error rate (see Figures~\ref{fig:2} and~\ref{fig:5}).

Finally, the assumption that $\sigma_i^2$ is known \citep{stephens2017false, fu2022heteroscedasticity} is often unrealistic in practice. When $\sigma_i^2$ is unknown, a common and natural approach is to substitute it with the sample variance $S_i^2$ and assume that $X_i \mid \mu_i, S_i^2 \sim N(\mu_i, S_i^2)$. However, this normal approximation can be inaccurate when $S_i^2$ substantially deviates from $\sigma_i^2$. To address this issue, \cite{stephens2017false} propose using a location-scale $t$-distribution for the conditional distribution of $X_i$ given $\mu_i$ and $S_i^2$. Nonetheless, this approach also faces similar challenges (see Figures~\ref{fig:E1}--\ref{fig:E6} in the Supplementary Material).

\subsection{Our Contributions} 
To address the limitations of existing methods, we propose a method called gg-Mix. Specifically, gg-Mix is an empirical Bayes method developed under the prior assumptions that will be formally introduced in \eqref{eqn:assum_prior_1} and \eqref{eqn:assum_prior_2}. Importantly, gg-Mix does not impose any structural constraints on $G$ or $f$; both are allowed to be arbitrary distributions supported on their respective domains. For further details on gg-Mix, see Section~\ref{sec2}.

We demonstrate the effectiveness of gg-Mix through extensive numerical studies and three real data examples, concluding that our method outperforms existing methods in the following aspects:
\begin{enumerate}
    \item gg-Mix adopts the empirical Bayes framework and can thereby incorporate structural information about the non-null distribution $f$, unlike methods developed under the empirical partially Bayes framework. To assess the effect of accounting for the structural information in $f$, we conduct numerical studies and find that gg-Mix outperforms empirical partially Bayes methods in terms of power, particularly when the structure of $f$ is highly informative. This aligns with the findings of \cite{sun2007oracle}.
    
    \item In the context of multiple testing, gg-Mix performs comparably to, and often outperforms, existing empirical Bayes methods. Specifically, the validity of existing methods heavily depends on the correctness of their prior assumptions, such as the inverse chi-squared assumption for $\sigma_i^2$ or the unimodality assumption for $f$. However, gg-Mix controls the type I error at the target level regardless of the underlying structure of $G$ and $f$, and achieves comparable or higher power than all other valid methods. This advantage arises from the flexible prior assumptions of gg-Mix, which can accommodate any structure for $G$ and $f$.

    \item gg-Mix does not require the true value of $\sigma_i^2$, and its validity does not depend on the discrepancy between $\sigma_i^2$ and its estimate $S_i^2$. This property is important in practice, since for degrees of freedom commonly encountered in applications, $\sigma_i^2$ may differ substantially from $S_i^2$.

    \item gg-Mix provides conservative estimates of the null proportion while remaining close to the true value when the null proportion is large (e.g., $\pi_0\geq0.7$), a setting commonly encountered in large-scale multiple testing problems. This balance is important because conservative estimation helps maintain the validity of multiple testing procedures, whereas excessive conservativeness can lead to a substantial loss of power.
\end{enumerate}

We conclude this section by highlighting the key distinctions between gg-Mix and MixTwice \citep{zheng2021mixtwice}, the existing method closely related to our approach. Both methods are developed under the empirical Bayes framework, assume independence between $\mu_i$ and $\sigma_i^2$, and adopt a nonparametric prior for $\sigma_i^2$. However, they differ in several important aspects.
First, MixTwice represents the non-null distribution of $\mu_i$ using discrete point masses, whereas gg-Mix models it using a continuous density. Second, the unimodality constraints in the two methods have different meanings: MixTwice imposes unimodality with mode at zero on the entire distribution of $\mu_i$; its non-null distribution is also unimodal but has its mode at a grid point adjacent to zero. In contrast, when using a unimodal working model, gg-Mix imposes unimodality with mode exactly at zero directly on the non-null distribution. Moreover, gg-Mix can relax this restriction by incorporating Gaussian location components, allowing it to accommodate non-unimodal shapes. Finally, the two methods differ in estimation: gg-Mix incorporates regularization to guard against underestimation of the null proportion, whereas MixTwice does not. Therefore, gg-Mix cannot be viewed simply as MixTwice with the unimodality constraint removed. These differences have important implications for null proportion estimation, FDR control, and power, as demonstrated in our numerical studies.

\medskip
\paragraph{Outline.} The remainder of this paper is organized as follows. In Section \ref{sec2}, we outline the general procedure of gg-Mix and illustrate its practical implementation. In Section \ref{sec3}, we conduct extensive numerical studies to evaluate the performance of gg-Mix and compare it with existing methods. In Section~\ref{sec4}, we apply gg-Mix to three real data examples previously analyzed by \cite{ignatiadis2025empirical}. In Section~\ref{sec5}, we discuss some limitations of gg-Mix and conclude with directions for future research.

\section{New Proposal}\label{sec2}
In this section, we first introduce the problem setting (Section~\ref{sub_sec_2.1}). We then describe the oracle version of our method, which assumes that the true prior distributions of $\sigma_i^2$ and $\mu_i$ are known (Section~\ref{sub_sec_2.2}). Next, we present a data-driven procedure that emulates the oracle procedure (Section~\ref{sub_sec_2.3}), and finally, we illustrate the practical implementation of the data-driven procedure (Section~\ref{sub_sec_2.4}).

\subsection{Problem Setting}\label{sub_sec_2.1}
Assume that for each $i \in [m]$, the observed pair $(X_i, S_i^2)$ arises from the following hierarchical model:
\begin{align}\label{eqn:assum_prior_1}
    \sigma_i^2 &\overset{\mathrm{i.i.d.}}{\sim} G(\cdot),\\ \label{eqn:assum_prior_2}
    \mu_i \mid \sigma_i^2 &\overset{\mathrm{i.i.d.}}{\sim} (1-\pi)~ \delta_0(\cdot) + \pi ~f(\cdot),\\ \label{eqn:like_2}
    (X_i, S_i^2) \mid \mu_i, \sigma_i^2 &\overset{\mathrm{ind.}}{\sim} N(\mu_i, \sigma_i^2)\ \otimes\ \frac{\sigma_i^2}{\nu}\chi^2_\nu,
\end{align}
where $G$ denotes any distribution supported on $[0, \infty)$, and $f$ denotes any density supported on $\mathbb{R}$. Because $\pi$ and $f$ are fixed regardless of $\sigma_i^2$, this prior specification implies independence between $\sigma_i^2$ and $\mu_i$, i.e., $\sigma_i^2 \indep \mu_i$.
We emphasize that no structural restrictions are imposed on either $G$ or $f$.
Given the observed values of $(X_i, S_i^2)$ generated from the hierarchical model described in \eqref{eqn:assum_prior_1}--\eqref{eqn:like_2}, our goal is to develop an empirical Bayes multiple testing procedure for the normal mean inference problem described in \eqref{eqn:hypo}.

As performance metrics for multiple testing procedures, we use the false discovery rate (FDR) and the true positive rate (TPR), which serve as measures of type I error and power, respectively. The FDR is defined as the expected value of the false discovery proportion (FDP), and the TPR as the expected value of the true positive proportion (TPP). Let $ \theta_i $ denote the true state of the $ i $-th hypothesis: if the $ i $-th null hypothesis is true, then $ \theta_i = 0 $; otherwise, $ \theta_i = 1 $. Let $ \delta_i $ denote the decision for the $i$-th hypothesis: if the $i$-th null hypothesis is rejected, then $ \delta_i = 1 $; otherwise, $ \delta_i = 0 $. Then, the FDR and TPR are defined as follows:
\begin{align*}
    \mathrm{FDR} &= \mathbb{E}[\mathrm{FDP}],\quad 
    \mathrm{FDP} = \frac{\sum_{i \in [m]} (1-\theta_i)\cdot\delta_i}{\max\left\{\sum_{i = 1}^m \delta_i,\, 1\right\}},\\
    \mathrm{TPR} &= \mathbb{E}[\mathrm{TPP}],\quad 
    \mathrm{TPP} = \frac{\sum_{i \in [m]} \theta_i \cdot \delta_i}{\max\left\{\sum_{i = 1}^m \theta_i,\, 1\right\}}.
\end{align*}

\subsection{The Oracle Procedure}\label{sub_sec_2.2}
Once the prior distribution is specified, an empirical Bayes multiple testing procedure proceeds as follows: First, the prior distribution is estimated from the observed data, $\{(X_i, S_i^2)\}_{i \in [m]}$. Using this estimated prior, a ranking statistic is calculated for each hypothesis. Without loss of generality, we assume that smaller values of the ranking statistic indicate stronger evidence against the null hypothesis (as is the case for $p$-values and Lfdr). The hypotheses are then ordered from smallest to largest according to their ranking statistics, and those with ranking statistics less than or equal to a chosen threshold are rejected.

In the oracle case, however, there is no need to estimate the prior distribution from the data, as it is assumed to be known. Thus, the only remaining tasks are to specify a ranking statistic, compute it using the known prior distribution, and determine the rejection threshold. 

We adopt $\mathrm{Lfdr}(X_i, S_i^2)$—the posterior probability that $\mu_i = 0$ given $X_i$ and $S_i^2$—as the ranking statistic for the $i$-th hypothesis. Lemma~\ref{lemma1} provides a way to compute the Lfdr under the prior assumptions specified in \eqref{eqn:assum_prior_1} and \eqref{eqn:assum_prior_2}. The proof of Lemma~\ref{lemma1} is straightforward and thus omitted.

\begin{lemma}\label{lemma1}
    Under the assumptions \eqref{eqn:assum_prior_1} and \eqref{eqn:assum_prior_2}, the posterior distribution of $\mu_i$ is given by:
    \begin{equation*}
        p(\mu_i \mid X_i, S_i^2) = \frac{p(X_i \mid \mu_i, S_i^2) \cdot p(\mu_i)}{\int  p(X_i \mid \mu_i, S_i^2) \cdot p(\mu_i) ~d\mu_i}, \quad i \in [m],
    \end{equation*}
    where the conditional distribution of $ X_i $ given $ \mu_i $ and $ S_i^2 $ can be expressed as:
    \begin{equation*}
        p(X_i \mid \mu_i, S_i^2) = \frac{\int p(X_i \mid \mu_i, \sigma_i^2) \cdot p(S_i^2 \mid \sigma_i^2) ~dG(\sigma_i^2)}{\int p(S_i^2 \mid \sigma_i^2) ~dG(\sigma_i^2)}.
    \end{equation*}
\end{lemma}

A key remaining question is how to choose a threshold in a data-driven manner to ensure control of the FDR at the target level. 
To address this, we adopt the procedure proposed by \cite{sun2007oracle}. 
Specifically, we define
\begin{equation*}
    k^* = \max \left\{ 
        k \in [m] : 
        \frac{1}{k} \sum_{i=1}^k \mathrm{Lfdr}_{(i)} \leq \alpha 
    \right\},
\end{equation*}
where $\mathrm{Lfdr}_{(i)}$ denotes the $i$-th smallest value of 
$\mathrm{Lfdr}(X_i, S_i^2)$, and $\alpha$ is the target FDR level. We adopt the convention that $\max \emptyset = 0$. Thus, if $k^*=0$, no hypotheses are rejected. Otherwise, we define the data-adaptive threshold as
\begin{equation*}
    \tau^* = \mathrm{Lfdr}_{(k^*)}.
\end{equation*}
Based on this threshold, the decision rule can be expressed as
\begin{equation*}
    \boldsymbol{\delta} = \left\{\,\delta_i = \mathbb{I}\left(\mathrm{Lfdr}(X_i, S_i^2) \leq \tau^*\right) : i \in [m]\,\right\},
\end{equation*}
where $\mathbb{I}(\cdot)$ denotes the indicator function.


\subsection{Data-Driven Procedure}\label{sub_sec_2.3}
In practice, however, the prior distribution is unknown. To address this, we emulate the oracle procedure using the plug-in principle: we estimate the unknown prior components $(G, f, \pi)$ from the observed data $ {(X_i, S_i^2)}_{i \in [m]} $, and substitute the estimated values for the unknown components in the procedure.

\subsubsection{Approximation for $G$ and $f$}\label{sub_sub_sec_2.3.1}
Since no structural restrictions are imposed on $G$ or $f$, their estimation is an infinite-dimensional problem and is therefore challenging in practice. To resolve this difficulty, we consider proxies for $G$ and $f$ that are more tractable while still capable of approximating the true priors with high accuracy. 

First, we approximate $G$ by $\tilde{G}$, a discretized version of the distribution supported on $L$ positive points:
\begin{align}\label{eqn:mix_g}
    \tilde{G} = \sum_{l=1}^L \Delta_l \cdot \delta_{\kappa_l},
\end{align}
where $\kappa_l$ and $L$ are user-specified values, and $\Delta_l$ are non-negative unknown mixing proportions that sum to one. 
This type of approximation has already been considered by \cite{zheng2021mixtwice} and \cite{ignatiadis2025empirical} as an alternative to the scaled inverse-chi-squared assumption, which is a conjugate prior and has been widely used in the literature \citep{baldi2001bayesian, lonnstedt2002replicated, smyth2004linear}.
Our numerical study in Section~\ref{sec3} demonstrates that approximating $G$ with a discrete distribution, as in \eqref{eqn:mix_g}, performs well regardless of whether the scaled inverse-chi-squared assumption holds.

Second, we approximate $f$ by $\tilde{f}$, using a finite mixture of continuous basis functions. Specifically, we consider two types of basis functions for constructing $\tilde{f}$:
\begin{align}\label{eqn:mix_f}
    U(\cdot ; \alpha, \beta) \quad \text{and} \quad N(\cdot ; \gamma, \zeta^2),
\end{align}
where $U(\cdot ; \alpha, \beta)$ denotes the uniform distribution supported on $(\alpha, \beta)$.
Three types of finite mixtures of the basis functions in \eqref{eqn:mix_f} that satisfy the unimodality assumption were previously introduced by \cite{stephens2017false}:
\begin{itemize}
    \item \textbf{Gaussian scale mixture:} $\tilde{f}(\cdot) = \sum_{k=1}^{K} \pi_k' N(\cdot; 0, \zeta_k^2)$
    \item \textbf{Uniform mixture:} $\tilde{f}(\cdot) = \sum_{k=1}^{K} \pi_k' U(\cdot; -\alpha_k, \alpha_k)$
    \item \textbf{Half-uniform mixture:} $\tilde{f}(\cdot) = \sum_{k=1}^{K/2} \pi_k' U(-\alpha_k, 0) + \sum_{k=K/2+1}^{K} \pi_k' U(0, \alpha_k)$
\end{itemize}
Here, $K$, $\alpha_k$, and $\zeta_k$ are user-specified values, while $\pi_k^\prime$ are non-negative unknown mixing proportions that sum to one. However, when the unimodality assumption is violated, none of these mixtures can adequately approximate $f$. To overcome this limitation, we propose an expanded mixture framework that augments the three unimodal mixture types introduced by \cite{stephens2017false} with an additional Gaussian location mixture component:
\begin{itemize}
    \item \textbf{Gaussian location + Gaussian scale mixture:}
    \[
    \tilde{f}(\cdot) = \sum_{k=1}^{K_1} \pi_k' N(\cdot; \gamma_k, \zeta^2) + \sum_{k=K_1+1}^{K_1 + K_2} \pi_k' N(\cdot; 0, \zeta_k^2)
    \]
    \item \textbf{Gaussian location + Uniform mixture:}
    \[
    \tilde{f}(\cdot) = \sum_{k=1}^{K_1} \pi_k' N(\cdot; \gamma_k, \zeta^2) + \sum_{k=K_1 + 1}^{K_1 + K_2} \pi_k' U(\cdot; -\alpha_k, \alpha_k)
    \]
    \item \textbf{Gaussian location + Half-uniform mixture:}
    \[
    \tilde{f}(\cdot) = \sum_{k=1}^{K_1} \pi_k' N(\cdot; \gamma_k, \zeta^2) + \sum_{k=K_1 + 1}^{K_1 + K_2/2} \pi_k' U(-\alpha_k, 0) + \sum_{k=K_1 + K_2/2 + 1}^{K_1 + K_2} \pi_k' U(0, \alpha_k)
    \]
\end{itemize}
Here, $K_1$, $K_2$, $\alpha_k$, $\gamma_k$, $\zeta$, and $\zeta_k$ are user-specified values, while $\pi_k^\prime$ are non-negative unknown mixing proportions that sum to one. 
For consistency, let $K = K_1 + K_2$. Then, there are six possible $K$-component mixtures that can be used to approximate $f$, and one can choose among them depending on the problem at hand.

\begin{remark}
The six working models for $f$ are nested in terms of their flexibility. With a sufficiently large and dense grid, the uniform mixture can approximate distributions that are symmetric and unimodal about zero, and the half-uniform mixture can approximate distributions that are unimodal about zero, with 
arbitrary accuracy. These approximation properties are supported by Khintchine's representation theorem \citep{khintchine1938unimodal}. Thus, in terms of the classes of distributions they can approximate, the following nested relationship holds:
\[
    \{\text{Gaussian scale mixture}\}
    \subset
    \{\text{uniform mixture}\}
    \subset
    \{\text{half-uniform mixture}\}.
\]
The same nesting relationship extends to the three extended working models obtained by adding a common Gaussian location mixture component.
\end{remark}

\begin{remark}\label{remark:choice_working_prior}
Despite this nested structure, we provide all six working models as options because the most flexible model is not necessarily the best choice. 
When reliable prior information about $f$ is available, we recommend using the simplest model consistent with that information, as this can improve estimation efficiency and reduce computational cost. In practice, however, reliable prior information about $f$ may not be available. In this case, choosing a simple working model entails the risk of misspecification, which may adversely affect both null proportion estimation and FDR control. A more flexible model is therefore a safer choice, as it reduces this risk. Even when a simpler model would have been sufficient, the main costs of using a more flexible model are some loss of estimation efficiency and additional computational cost.
\end{remark}

\subsubsection{Estimation of Unknown Parameters}
Under the approximations of $G$ and $f$ described in Section~\ref{sub_sub_sec_2.3.1}, the unknown parameters include the mixing proportions $\{\Delta_l\}_{l = 1}^L$ associated with $\tilde{G}$, the mixing proportions $\{\pi_k^\prime\}_{k = 1}^K$ associated with $\tilde{f}$, and the null proportion $\pi_0 \coloneqq 1 - \pi$. In total, there are $L + K + 1$ unknown parameters to be estimated.

Our estimation strategy is two-fold. First, we estimate the mixing proportions $ \{\Delta_l\}_{l = 1}^L $ by maximizing an approximation of the marginal likelihood of $ S_i^2 $. Then, conditional on these estimates, we estimate the mixing proportions $ \{\pi_k^\prime\}_{k = 1}^K $ and the null proportion $ 1 - \pi $ by maximizing an approximation of the regularized profile conditional likelihood of $ X_i$ given $S_i^2 $ as in \cite{stephens2017false}.
 
To obtain an estimate of $\boldsymbol{\Delta} = \{\Delta_l : l \in [L]\}$, we consider an approximation for the marginal likelihood of 
$S_i^2$, denoted by $p(S_i^2 ; \boldsymbol{\Delta})$,
\begin{align*}
    S_i^2 \sim \int p(S_i^2 \mid \sigma_i^2) ~dG(\sigma_i^2) 
    \approx \int p(S_i^2 \mid \sigma_i^2) ~d\tilde{G}(\sigma_i^2; \boldsymbol{\Delta})
    \eqqcolon p(S_i^2; \boldsymbol{\Delta}).
\end{align*}
By solving the optimization problem
\begin{align}\label{opt1}
    \hat{\boldsymbol{\Delta}} = \underset{\boldsymbol{\Delta}}{\arg\max} \sum_{i = 1}^m \log p(S_i^2 ; \boldsymbol{\Delta}),
\end{align}
we can obtain an estimate of $\boldsymbol{\Delta}$. The objective function in \eqref{opt1} is concave with respect to $\boldsymbol{\Delta}$, and we solve it by applying the sequential quadratic programming method proposed by \cite{kim2020fast}.

The remaining unknown parameters are the null proportion $1 - \pi$ and the mixing proportions $\{\pi_k^\prime\}_{k = 1}^K$. Let $\boldsymbol{\pi} = \{\pi_0, \pi_1, \ldots, \pi_K\}$, where $\pi_0 = 1 - \pi$ and $\pi_k = (1-\pi_0) \cdot \pi_k^\prime$ for all $k \in [K]$. 
To obtain an estimate of $\boldsymbol{\pi}$, we consider the approximation of the conditional distribution of $X_i$ given $S_i^2$, denoted by $p(X_i \mid S_i^2; \boldsymbol{\Delta}, \boldsymbol{\pi})$: 
\begin{align*}
    X_i \mid S_i^2 
    &\sim \int p(X_i \mid \mu_i, S_i^2) \cdot p(\mu_i) ~d\mu_i\\
    &= \int p(X_i \mid \mu_i, S_i^2) \cdot \pi_0 ~\delta_0(d\mu_i) + \int p(X_i \mid \mu_i, S_i^2) \cdot (1-\pi_0)  ~f(\mu_i)~d\mu_i\\
    &\approx \int p(X_i \mid \mu_i, S_i^2; \boldsymbol{\Delta}) \cdot \pi_0 ~ \delta_0(d\mu_i) + \int p(X_i \mid \mu_i, S_i^2; \boldsymbol{\Delta}) \cdot (1-\pi_0) ~\tilde{f}(\mu_i; \pi_k^\prime, k \in [K]) ~d\mu_i\\
    &\eqqcolon p(X_i \mid S_i^2; \boldsymbol{\Delta}, \boldsymbol{\pi}),
\end{align*}
where
\begin{align*}
    p(X_i \mid \mu_i, S_i^2; \boldsymbol{\Delta}) &\coloneqq \int p(X_i \mid \mu_i, \sigma_i^2) \cdot p(\sigma_i^2 \mid S_i^2; \boldsymbol{\Delta}) ~d\sigma_i^2, \quad \text{and} \\
    p(\sigma_i^2 \mid S_i^2; \boldsymbol{\Delta}) &\coloneqq \dfrac{p(S_i^2 \mid \sigma_i^2)  ~d\tilde{G}(\sigma_i^2; \boldsymbol{\Delta})}{\int p(S_i^2 \mid \sigma_i^2) ~d\tilde{G}(\sigma_i^2; \boldsymbol{\Delta})} .
\end{align*}
Given the estimate of $ \hat{\boldsymbol{\Delta}} = \{\hat{\Delta}_l : l \in [L]\} $, we solve the following optimization problem to obtain an estimator for $\boldsymbol{\pi}$: 
\begin{align}\label{opt2}
    \hat{\boldsymbol{\pi}}(\lambda) = \underset{\boldsymbol{\pi}}{\arg\max} \sum_{i = 1}^m \log p(X_i \mid S_i^2; \hat{\boldsymbol{\Delta}}, \boldsymbol{\pi}) + (\lambda-1) \log(\pi_0),
\end{align}
where $\lambda \geq 1$ is a penalty parameter that controls the regularization strength, with larger values imposing stronger regularization to prevent underestimation of the null proportion.
Given $\lambda$, the objective function in \eqref{opt2} is concave with respect to $\boldsymbol{\pi}$, and we solve the optimization problem using the sequential quadratic programming method of \cite{kim2020fast}.

\subsubsection{Data-driven Decision Rule}
We consider the following approximation of $\mathrm{Lfdr}(X_i, S_i^2)$:
\begin{align*}
    \text{Lfdr}(X_i, S_i^2) 
    &= p(\mu_i = 0 \mid X_i, S_i^2)\\
    &= \frac{p(X_i \mid \mu_i = 0, S_i^2) \cdot p(\mu_i = 0)}{\int p(X_i \mid \mu_i, S_i^2) \cdot p(\mu_i) ~d\mu_i}\\
    &\approx \dfrac{p(X_i \mid \mu_i = 0, S_i^2; \boldsymbol{\Delta}) \cdot p(\mu_i = 0)}{\int p(X_i\mid \mu_i, S_i^2; \boldsymbol{\Delta}) \cdot p(\mu_i ; \boldsymbol{\pi})d\mu_i }\\
    &\eqqcolon \text{Lfdr}(X_i, S_i^2; \boldsymbol{\Delta}, \boldsymbol{\pi}),
\end{align*}
where 
\begin{align*}
    p(\mu_i ; \boldsymbol{\pi}) &\coloneqq \pi_0 \cdot \delta_0(\mu_i) + (1-\pi_0) \cdot \tilde{f}(\mu_i ; \pi_k^\prime, k \in [K]).
\end{align*}
By plugging in the unknown parameters $\boldsymbol{\Delta}$ and $\boldsymbol{\pi}$ with the solutions to the optimization problems described in \eqref{opt1} and \eqref{opt2}, we obtain the estimated Lfdr values and define
\begin{align*}
    \hat{k} = \max \left\{
        k \in [m] :
        \frac{1}{k} \sum_{i=1}^k \widehat{\mathrm{Lfdr}}_{(i)} \leq \alpha
    \right\},
\end{align*}
where $\widehat{\mathrm{Lfdr}}_{(i)}$ denotes the $i$-th smallest value of
$\mathrm{Lfdr}(X_i, S_i^2; \hat{\boldsymbol{\Delta}}, \hat{\boldsymbol{\pi}})$,
and $\alpha$ is the target FDR level.
We adopt the convention that $\max \emptyset = 0$. Thus, if $\hat{k}=0$, no hypotheses are rejected. Otherwise, we define
the data-adaptive threshold as
\begin{equation}\label{opt_thr}
    \hat{\tau} = \widehat{\mathrm{Lfdr}}_{(\hat{k})}.
\end{equation}
Based on this threshold, we construct the decision rule
\begin{align}\label{dec_rule}
    \boldsymbol{\delta}
    =
    \left\{
        \delta_i =
        \mathbb{I}\left(
        \mathrm{Lfdr}\left(
        X_i, S_i^2;
        \hat{\boldsymbol{\Delta}}, \hat{\boldsymbol{\pi}}
        \right)
        \leq \hat{\tau}
        \right)
        : i \in [m]
    \right\}.
\end{align}
Algorithm~\ref{alg_1} summarizes our proposed method:

\begin{algorithm}
    \caption{gg-Mix}\label{alg_1}
    \textbf{Input:} $\{(X_i,S_i^2)\}_{i\in[m]}$, user-specified values for $\tilde{G}$ and $\tilde{f}$, target FDR level $\alpha$, and penalty parameter $\lambda$.
    
    \textbf{Output:} The set of rejected indices $\mathcal{R}  = \{i \in [m] : \delta_i = 1\}$.
    \begin{algorithmic}[1]     
        \State Estimate $\boldsymbol{\Delta}$ by solving the optimization problem defined in \eqref{opt1}.

        \State Estimate $\boldsymbol{\pi}$ by solving the optimization problem defined in \eqref{opt2}.
    
        \State Calculate the threshold $\hat{\tau}$ using the formula in \eqref{opt_thr}, and then construct the decision rule as in \eqref{dec_rule}.
    
    \end{algorithmic}
    \textbf{Return:} the set of rejected indices $\mathcal{R}$. 
\end{algorithm}

\subsection{Implementation Details}\label{sub_sec_2.4}
In this section, we discuss the practical implementation details of gg-Mix, including the choice of user-specified values for $\tilde{G}$ and $\tilde{f}$, as well as the penalty parameter $\lambda$.

To specify $\tilde{G}$ in \eqref{eqn:mix_g}, we need to determine the number of mixture components $L$ and the corresponding grid points $\{\kappa_l\}_{l=1}^L$. We recommend choosing a sufficiently large value for $L$, such as $L \geq 50$, to ensure adequate flexibility in capturing the marginal distribution of $S_i^2$. 
Given a fixed $L$, we set $a$ to the 1\% quantile and $b$ to the maximum of the observed $S_i^2$ values. We then generate $L$ logarithmically equally spaced grid points between $a$ and $b$, denoted by $\kappa_1<\cdots<\kappa_L$, with $\kappa_1=a$ and $\kappa_L=b$. 
A sensitivity analysis with respect to $L$ indicates that $L=50$ is sufficient for all choices of $G$ considered in this paper and that the performance of gg-Mix is insensitive to moderate changes in $L$. See Section~\ref{appendix_b} of the Supplementary Material for details.

Now, consider the specification of $\tilde{f}$. Recall that there are six candidate mixture distributions for modeling $\tilde{f}$ (see Section~\ref{sub_sub_sec_2.3.1}). To implement $\tilde{f}$ in practice, one must first choose a working model from among these candidates and then specify the corresponding user-specified values. As already discussed in Remark~\ref{remark:choice_working_prior}, when there is no strong prior belief about the true density $f$, we recommend selecting one of the extended mixture models that includes an additional Gaussian location mixture component for greater flexibility.
In this paper, for all simulation settings, we adopt the Gaussian location $+$ Gaussian scale mixture as the working model for $\tilde{f}$:
\begin{equation*}
\tilde{f}(\cdot) = \sum_{k=1}^{K_1} \pi_k' N(\cdot; \gamma_k, \zeta^2) + \sum_{k=K_1+1}^{K_1 + K_2} \pi_k' N(\cdot; 0, \zeta_k^2).
\end{equation*}
This choice allows $\tilde{f}$ to flexibly accommodate a wide range of non-null distributions, including those that deviate from unimodality.

To implement the Gaussian location mixture part, the user must specify the number of components $K_1$, the grid of location parameters $\{\gamma_k\}_{k=1}^{K_1}$, and the common variance $\zeta^2$. We recommend setting $K_1$ to a sufficiently large value (e.g., $K_1 \geq 50$) to ensure adequate modeling flexibility. Given a fixed $K_1$, we set $a$ and $b$ to the 1\% lower and upper quantiles of the observed $X_i$ values, respectively. We then construct $K_1$ equally spaced grid points between $a$ and $b$, denoted by $\gamma_1<\cdots<\gamma_{K_1}$, with $\gamma_1=a$ and $\gamma_{K_1}=b$. By default, we fix the variance $\zeta^2$ at 1. A sensitivity analysis with respect to $K_1$ indicates that $K_1=50$ is sufficient for all choices of $f$ considered in this paper and that the performance of gg-Mix is insensitive to moderate changes in $K_1$. See Section~\ref{appendix_b} of the Supplementary Material for details.

For the Gaussian scale mixture part, the user must specify the number of components $K_2$ and the component-specific variances $\{\zeta_k^2\}_{k = K_1 + 1}^{K_1 + K_2}$. We follow the recommendations of \cite{stephens2017false}, which have been shown to work well in a variety of simulation settings. For detailed guidelines and the specific grid construction, we refer readers to the supplementary material of \cite{stephens2017false}.

The penalty parameter $\lambda$ in \eqref{opt2} is related to the null proportion. Choosing a large value of 
$\lambda$ makes the estimated null proportion more conservative, closer to 1, while a small value results in a more liberal estimate. \cite{stephens2017false} recommends using 10 as a default value, and we adopt it. While this value may not be optimal, it performs well across all our simulation settings.
A sensitivity analysis with respect to $\lambda$ shows that the penalty helps guard against underestimation of the null proportion and that the performance of gg-Mix is insensitive to moderate changes in $\lambda$, provided that $\lambda$ is not too small. See Section~\ref{appendix_c} of the Supplementary Material for details.

\section{Numerical Study}\label{sec3}

In this section, we compare the performance of gg-Mix with existing methods using simulation data generated from the hierarchical model described in \eqref{eqn:assum_prior_1}--\eqref{eqn:like_2}. We also assess the robustness of gg-Mix to violations of the independence assumptions, considering dependence between $\mu_i$ and $\sigma_i^2$ as well as dependence across the pairs $\{(X_i,S_i^2)\}_{i=1}^m$. The results of these additional simulation studies are provided in Section~\ref{appendix_f} of the Supplementary Material.

Throughout all simulation settings, we fix the number of hypotheses at $m = 5,000$. 
Under the prior assumptions in \eqref{eqn:assum_prior_1} and \eqref{eqn:assum_prior_2}, we need to specify three components: $G$, $f$, and $\pi_0$. For $G$, we consider the following three distributions:
\begin{enumerate}
\item[(g1)] Scaled inverse-chi-squared: $\mathrm{Inv}\text{-}\chi^2_{6} / 6$;
\item[(g2)] Point mass at $1$: $\delta_1$;
\item[(g3)] Two-point mass assigning equal weight to $1$ and $10$: $0.5 \delta_1 + 0.5 \delta_{10}$.
\end{enumerate}
For $f$, we consider the following three density functions:
\begin{enumerate}
\item[(f1)] Gaussian distribution: $N(0, 4^2)$;
\item[(f2)] Gaussian scale mixture: $\frac{2}{3} N(0, 1) + \frac{1}{3} N(0, 2^2)$;
\item[(f3)] Gaussian location mixture: $\pi_1' N(-3, 1) + (1 - \pi_1') N(3, 1)$, where $\pi_1' \in [0.5, 1]$.
\end{enumerate}
For $\pi_0$, we consider the following nine levels: $0.1, 0.2, \ldots, 0.9$.

Under the likelihood assumption in \eqref{eqn:like_2}, we need to specify the degrees of freedom $\nu$. We consider the following six levels of $\nu$: $2, 4, 8, 16, 32, 64$.

To evaluate the performance of each method, we consider three metrics: (i) the estimation accuracy of the null proportion, assessed by comparing the true value to the average of the estimated null proportions over 200 replications; (ii) the type I error, measured by the (empirical) FDR, computed as the average FDP across 200 replications; and (iii) power, measured by the (empirical) TPR, computed as the average TPP over 200 replications. The target FDR level is set to $\alpha = 0.1$ across all simulation settings.

\subsection{Comparing Methods}
For comparison, we consider the following four methods.

\begin{enumerate}
    \item[(a)] IS: IS denotes the method proposed by \cite{ignatiadis2025empirical}, which is developed under the empirical partially Bayes framework. In this approach, the prior distribution $G$ for $\sigma_i^2$ is modeled nonparametrically, and is estimated using a nonparametric maximum likelihood estimator (NPMLE). Once $G$ is estimated, conditional $p$-values are computed based on these estimates, and the BH procedure is subsequently applied to obtain the rejection set. In our implementation, we follow the practical recommendations outlined in \cite{ignatiadis2025empirical} for estimating the prior density $G$, which essentially involves solving the optimization problem presented in \eqref{opt1}. We solve this optimization problem using the sequential quadratic programming method proposed by \cite{kim2020fast}.

    \item[(b)] ash-TwoStep: ash-TwoStep denotes the method proposed by \cite{lu2019empirical}, which is developed under the empirical Bayes framework. This approach assumes the following prior distributions:
    \begin{equation*}
        \sigma_i^2 \overset{\mathrm{i.i.d.}}{\sim} \text{Scaled-Inv-}\chi^2, \quad \mu_i \overset{\mathrm{i.i.d.}}{\sim} (1-\pi)~ \delta_0(\cdot) + \pi~ f(\cdot),
    \end{equation*}
    where $f$ is a unimodal density supported on $\mathbb{R}$.
    In our implementation, we use the \texttt{ashr} package in R. Specifically, we select the half-uniform mixture option provided by \texttt{ashr} to model $f$. This option allows us to approximate an arbitrary unimodal density.

    \item[(c)] MixTwice: MixTwice denotes the method proposed by \cite{zheng2021mixtwice}, which is developed under the empirical Bayes framework. This approach can be viewed as a nonparametric extension of ash-TwoStep, as it replaces the scaled inverse-chi-squared prior with an arbitrary distribution function supported on $[0, \infty)$. Like \cite{lu2019empirical}, MixTwice also adopts the unimodality assumption for $f$. We implement this method using the \texttt{MixTwice} package in R.

    \item[(d)] gg-Mix: gg-Mix denotes our proposed method, which is developed under the empirical Bayes framework. This approach can be viewed as an extension of MixTwice in that it removes the unimodality assumption for $f$. For further implementation details of gg-Mix, see Section~\ref{sub_sec_2.4}.
\end{enumerate}

In addition to the aforementioned four methods, we also consider the methods proposed by \cite{benjamini1995controlling} and \cite{smyth2004linear}, as well as two alternative versions of the method by \cite{stephens2017false}: one based on normal approximation and the other on location-scale $t$ approximation. The performance of the methods by \cite{benjamini1995controlling} and \cite{smyth2004linear} is dominated by that of IS. The two versions by \cite{stephens2017false} fail to control the FDR at the target level in all but a few simulation settings. Therefore, we report their results in Section~\ref{appendix_e} of the Supplementary Material.

\subsection{Results}
\subsubsection{Effect of Null Proportion $\pi_0$}\label{sub_sub_sec_3.2.1}

\begin{figure}[htb!]
    \centering
    \includegraphics[width=0.95\linewidth]{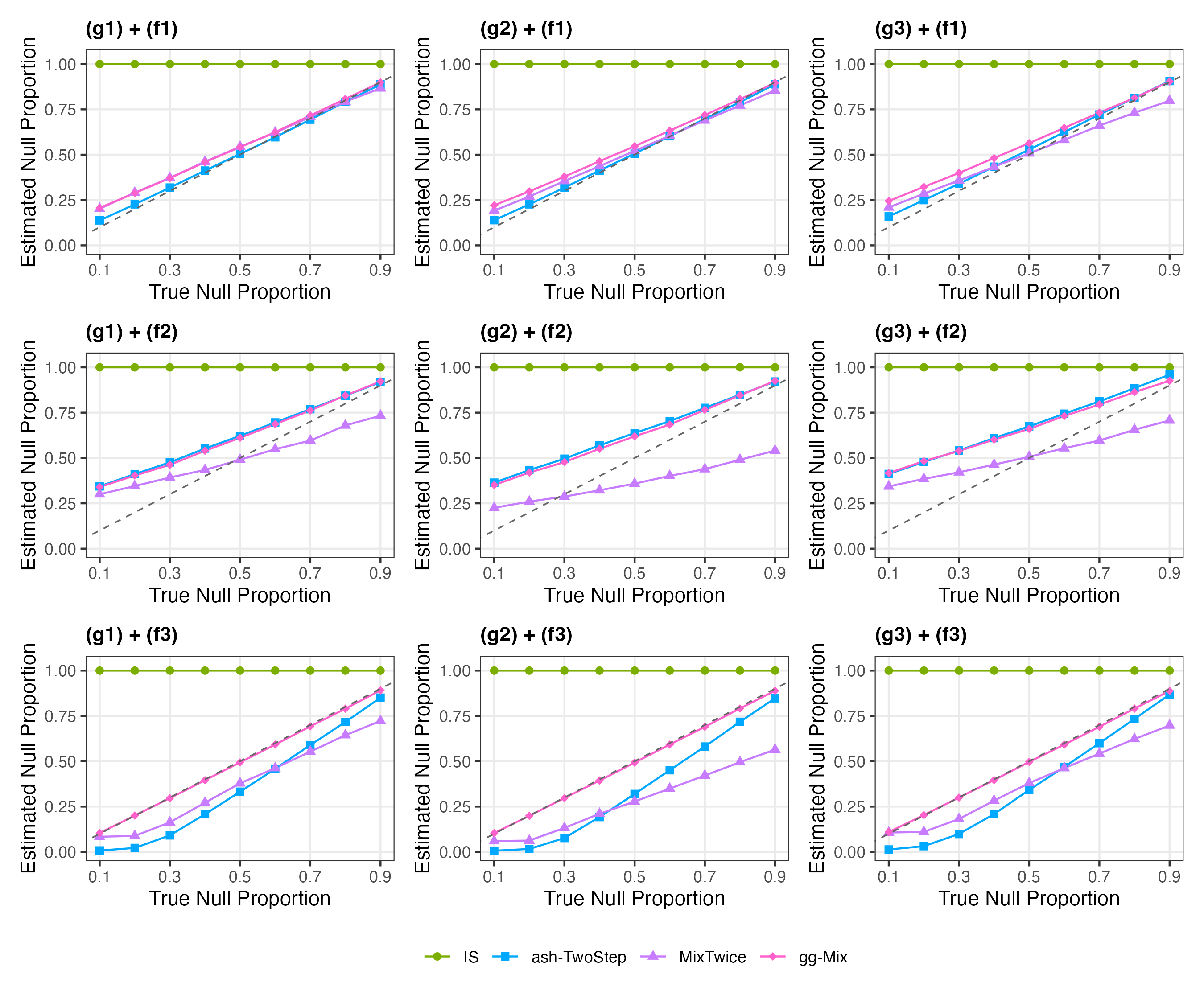}
    \caption{Estimated Null Proportion: Each column, from left to right, corresponds to the results for priors (g1) to (g3), while each row, from top to bottom, corresponds to priors (f1) to (f3). In each panel, the x-axis represents the true null proportion, and the y-axis shows the estimated null proportion. The black dashed lines indicate the 45-degree line.}


    \label{fig:1}
\end{figure}

\begin{figure}[htb!]
    \centering
    \includegraphics[width=0.95\linewidth]{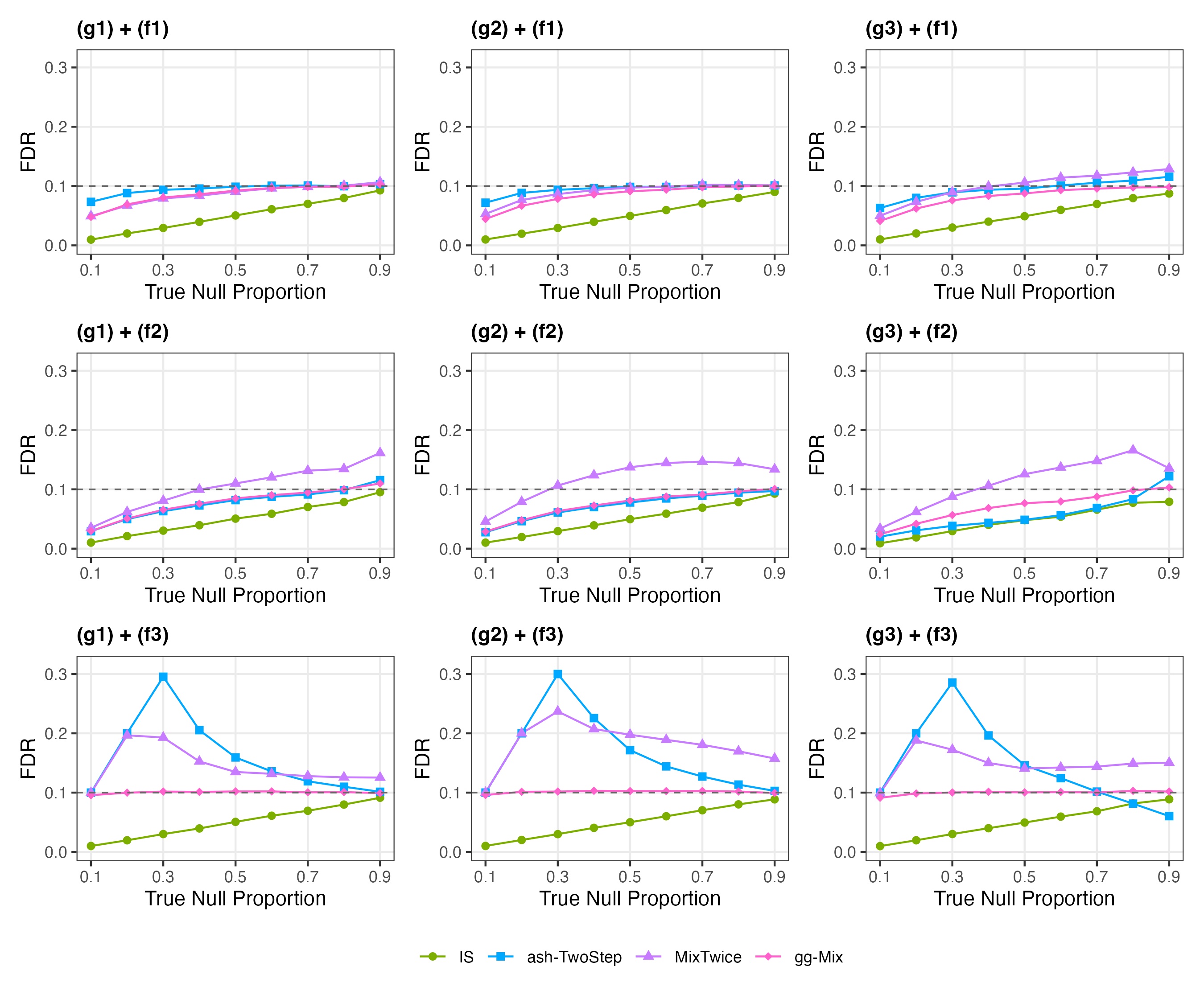}
    \caption{(Empirical) False Discovery Rate: Each column, from left to right, corresponds to the results for priors (g1) to (g3), while each row, from top to bottom, corresponds to priors (f1) to (f3). In each panel, the x-axis represents the null proportion, and the y-axis shows the FDR. The horizontal dashed lines represent the target FDR level of 0.1.}


    \label{fig:2}
\end{figure}

\begin{figure}[htb!]
    \centering
    \includegraphics[width=0.95\linewidth]{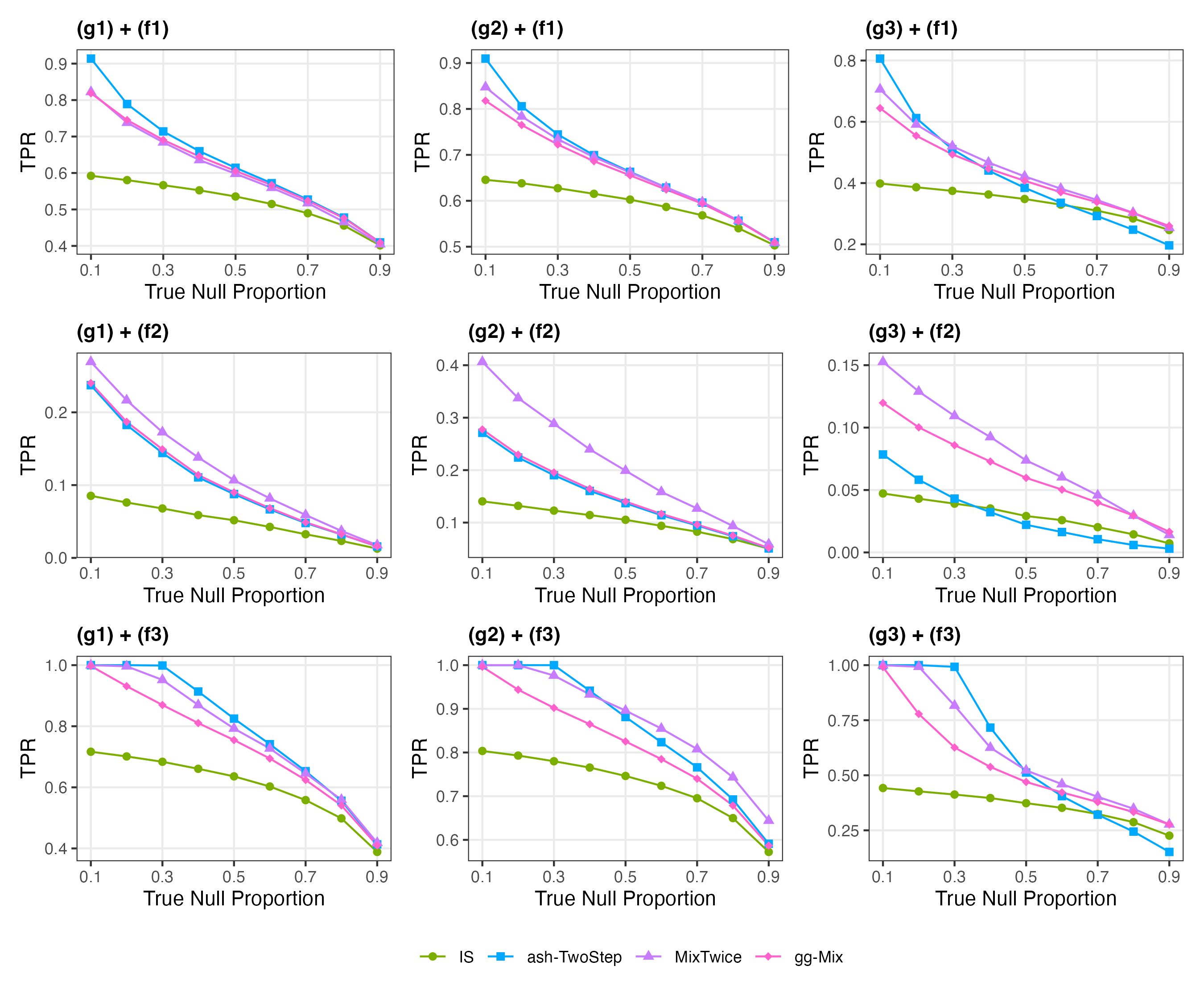}
    \caption{(Empirical) True Positive Rate: Each column, from left to right, corresponds to the results for priors (g1) to (g3), while each row, from top to bottom, corresponds to priors (f1) to (f3). In each panel, the x-axis represents the null proportion, and the y-axis shows the TPR.}

    \label{fig:3}
\end{figure}

This section is dedicated to investigating the effect of the null proportion $\pi_0$ on the performance of each method. To this end, we fix $\nu = 10$ and $\pi_1' = 0.5$, and consider all possible combinations of the three types of $G$, the three types of $f$, and the nine levels of $\pi_0$, which yields a total of $3 \times 3 \times 9 = 81$ simulation settings.
The results are summarized in Figures~\ref{fig:1}--\ref{fig:3}. From these results, we make the following observations.

First, both gg-Mix and IS successfully control the FDR at the target level of $0.1$ across all simulation settings. However, gg-Mix consistently achieves higher power than IS. This power loss of IS is partly attributable to its use of the BH procedure for multiplicity correction, which does not estimate the null proportion and effectively corresponds to using $\pi_0=1$. As a result, IS becomes increasingly conservative as the true null proportion decreases, leaving more of the available FDR budget unused and resulting in a substantial loss of power. In particular, when $\pi_0$ is close to $0.1$, the FDR of IS falls significantly below the target level, and the power gap between gg-Mix and IS becomes more pronounced.

Second, ash-TwoStep exhibits similar patterns in the estimated $\pi_0$, FDR, and TPR curves to gg-Mix when its prior assumptions---the scaled inverse-chi-squared and unimodality assumptions---are satisfied (see the panels corresponding to the first two columns and rows in each figure). In particular, when $f=(f1)$ and the true null proportion is close to $0.1$, ash-TwoStep achieves the most accurate estimate of the null proportion, makes the most efficient use of the given FDR budget, and consequently attains the highest power among all methods. However, misspecification of its prior assumptions can adversely affect both FDR control and power. For example, when the unimodality assumption is violated (i.e., $f=(f3)$), the FDR of ash-TwoStep substantially exceeds the target level across most scenarios, which is closely related to underestimation of the null proportion. 
Prior misspecification can also result in a substantial loss of power even when the FDR remains close to the target level. For example, under $(g3)+(f2)$, where the scaled inverse-chi-squared assumption is violated, ash-TwoStep maintains the FDR below or near the target level but consistently exhibits lower power than gg-Mix across all null proportions.

Third, MixTwice exhibits a similar pattern in estimated $\pi_0$, FDR, and TPR curves to gg-Mix when $f = (f1)$. However, when $f = (f2)$ or $f = (f3)$, MixTwice fails to control the FDR in some settings. For $f = (f3)$, this failure is expected since the unimodality assumption required by MixTwice is violated. In contrast, for $f = (f2)$, the prior assumption of MixTwice is satisfied, yet FDR control is not achieved in certain cases. In particular, when $f = (f2)$ and $G = (g2)$, MixTwice significantly underestimates the null proportion over the range of null proportions that are most likely to occur in practice (e.g., $\pi_0 \geq 0.7$), which results in a failure to control the FDR in this range.

Finally, under (f3), the FDRs of ash-TwoStep and MixTwice increase as $\pi_0$ decreases from $0.9$, showing substantial inflation around $\pi_0=0.2$--$0.3$, but decline toward the target level as $\pi_0$ approaches $0.1$. The initial increase reflects the increasing impact of violating the unimodality assumption as the non-null proportion increases. The subsequent decline may appear counterintuitive, but is explained by the small number of null hypotheses: when $\pi_0=0.1$, even rejecting all hypotheses yields an FDP of $0.1$.

\subsubsection{Effect of Degrees of Freedom $\nu$}
\begin{figure}[htb!]
    \centering
    \includegraphics[width=0.95\linewidth]{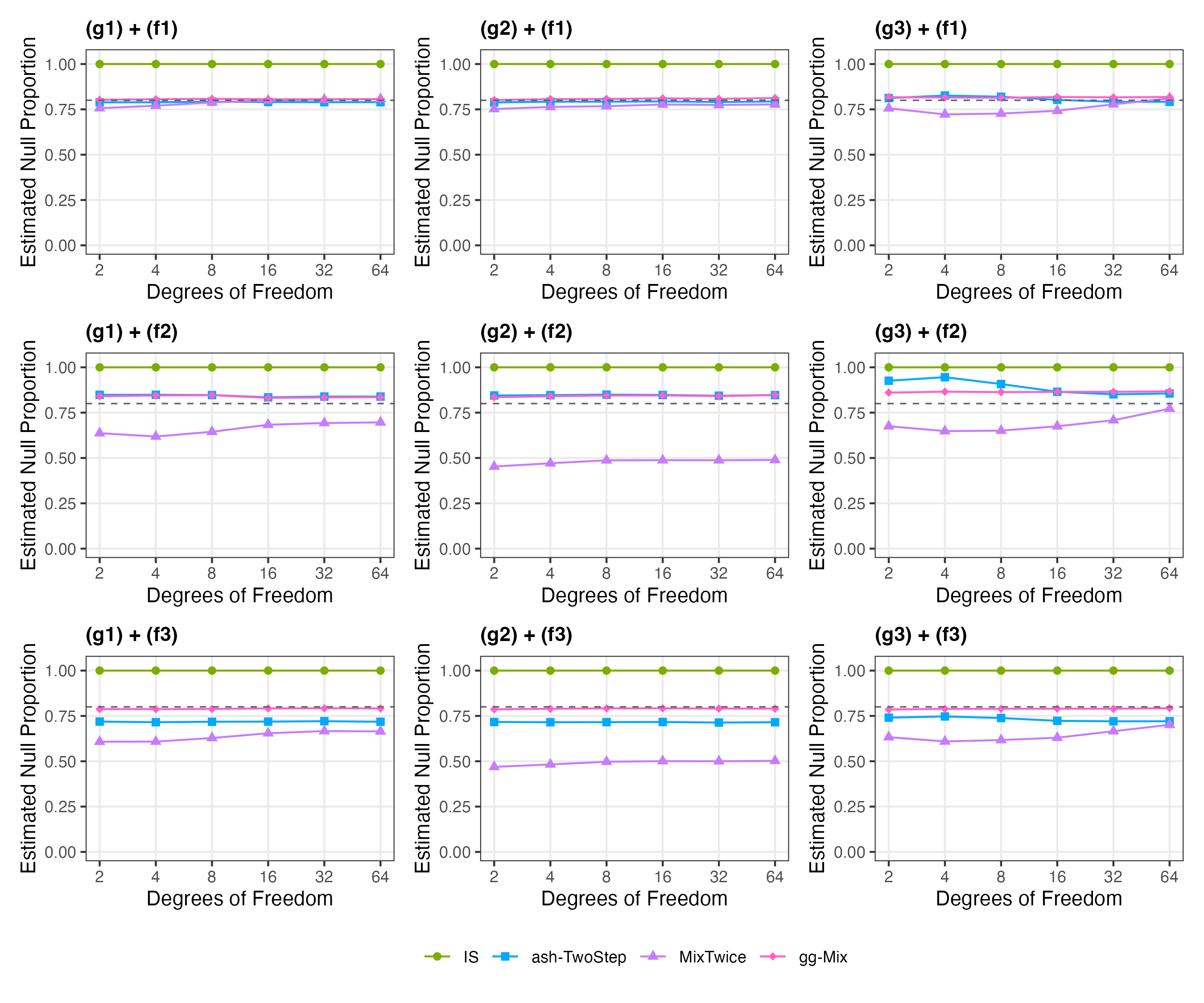}
    \caption{Estimated Null Proportion: Each column, from left to right, corresponds to the results for priors (g1) to (g3), while each row, from top to bottom, corresponds to priors (f1) to (f3). In each panel, the x-axis represents the degrees of freedom, and the y-axis shows the estimated null proportion. The black dashed line indicates the true null proportion of 0.8.}
    \label{fig:4}
\end{figure}

\begin{figure}[htb!]
    \centering
    \includegraphics[width=0.95\linewidth]{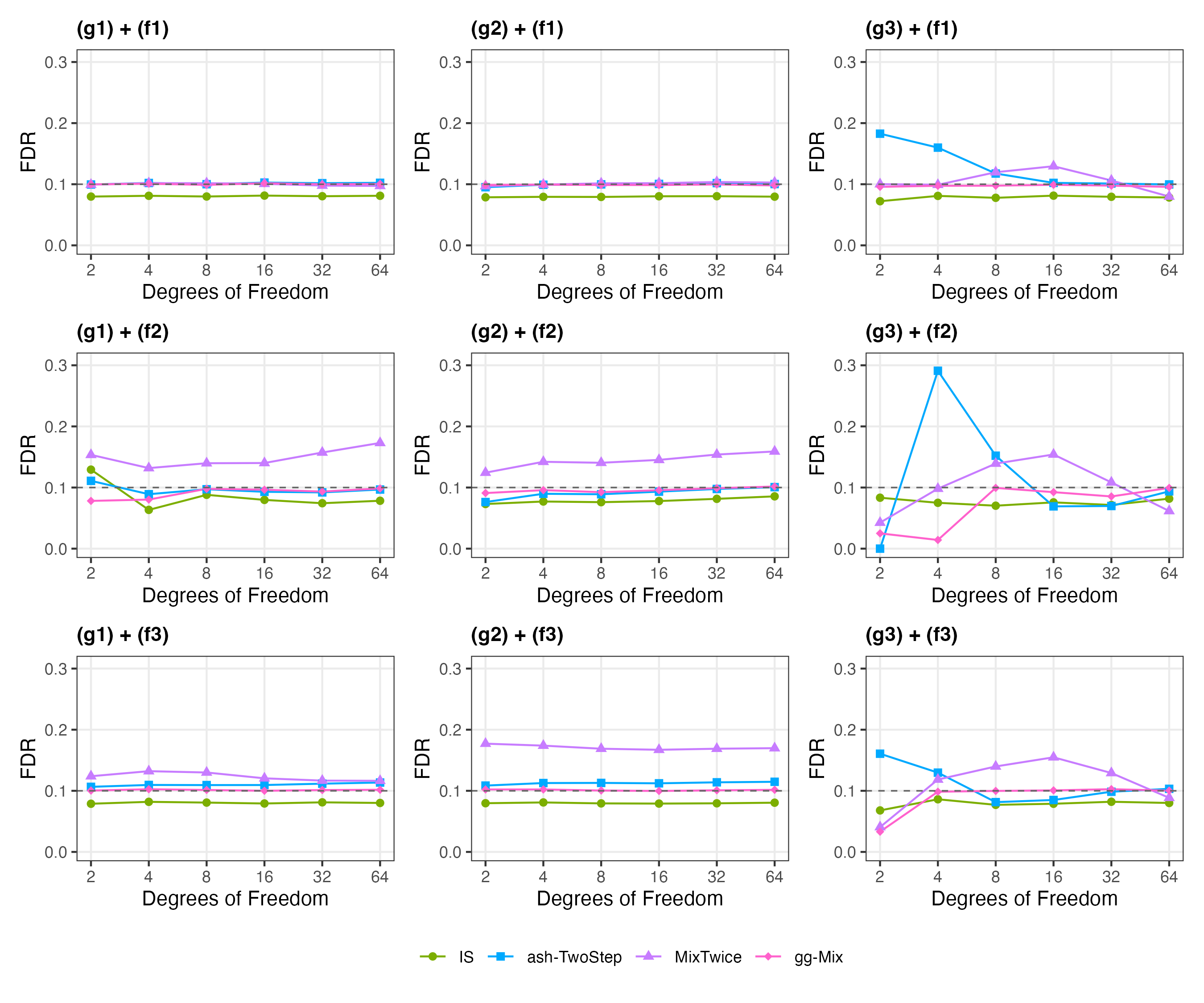}
    \caption{(Empirical) False Discovery Rate: Each column, from left to right, corresponds to the results for priors (g1) to (g3), while each row, from top to bottom, corresponds to priors (f1) to (f3). In each panel, the x-axis represents the degrees of freedom, and the y-axis shows the FDR. The horizontal dashed lines represent the target FDR level of 0.1.}
    \label{fig:5}
\end{figure}

\begin{figure}[htb!]
    \centering
    \includegraphics[width=0.95\linewidth]{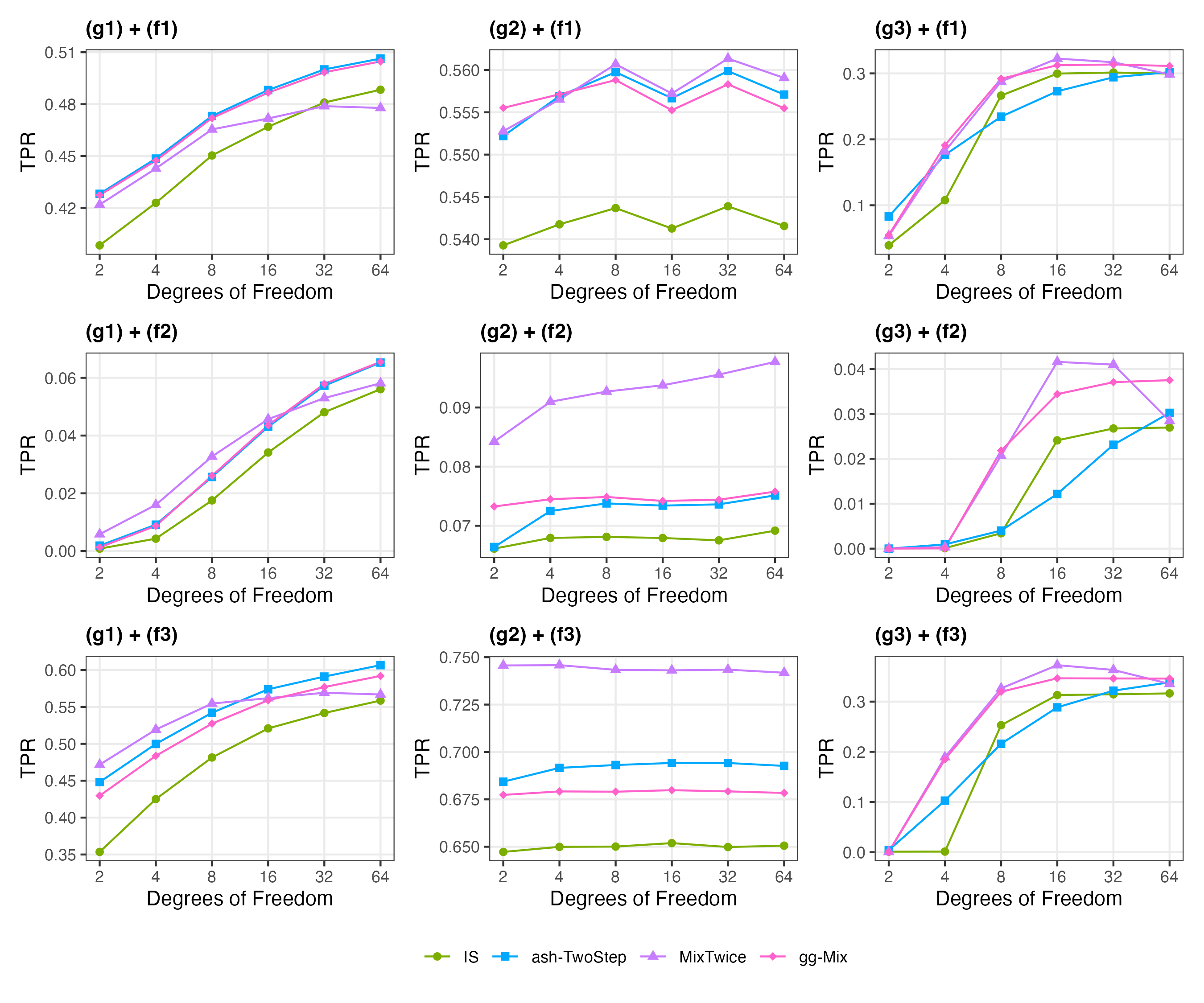}
    \caption{(Empirical) True Positive Rate: Each column, from left to right, corresponds to the results for priors (g1) to (g3), while each row, from top to bottom, corresponds to priors (f1) to (f3). In each panel, the x-axis represents the degrees of freedom, and the y-axis shows the TPR.}
    \label{fig:6}
\end{figure}

Now, we investigate the effect of degrees of freedom $\nu$ on the performance of each method. To this end, we fix $\pi_0 = 0.8$ and $\pi_1' = 0.2$, and consider all possible combinations of the three types of $G$, the three types of $f$, and six levels of $\nu$, which yields a total of $3 \times 3 \times 6 = 54$ simulation settings.
The results are summarized in Figures~\ref{fig:4}--\ref{fig:6}.

We find patterns similar to those presented in Section~\ref{sub_sub_sec_3.2.1}. gg-Mix controls the FDR regardless of the underlying prior distributions and degrees of freedom, while consistently achieving comparable or the highest power among all valid methods. IS also controls the FDR in all simulation settings. However, as mentioned before, IS always sets the null proportion to 1, which is the most conservative choice and leads to a loss of power. ash-TwoStep shows a similar pattern to gg-Mix when its prior assumptions are satisfied, but can exhibit FDR inflation or power loss when these assumptions are violated. 
In particular, when $G=(g3)$, the scaled inverse-chi-squared assumption for $G$ is violated, and ash-TwoStep exhibits substantial FDR inflation for some values of $\nu$, particularly at small degrees of freedom. As $\nu$ increases, its FDR approaches the target level, but its power remains lower than that of gg-Mix.
MixTwice also shows a similar pattern to gg-Mix when $f = (f1)$, but fails to control the FDR in some settings even when its prior assumptions are satisfied.

\subsubsection{Effect of Structural Information in the Non-null Distribution $f$ on Power}
In this section, we demonstrate the importance of incorporating the structural information of $f$ into the testing procedure. To this end, we fix $\pi_0 = 0.8$ and $\nu = 10$, and focus on the case where $f = (f3)$, considering varying levels of the mixing proportion $\pi_1^\prime$. Specifically, we consider all possible combinations of the three types of $G$ and six levels of $\pi_1^\prime$, ranging from $0.5$ to $1$ in increments of $0.1$. This yields a total of $3 \times 6 = 18$ simulation settings. 

\begin{figure}[htb!]
    \centering
    \includegraphics[width=0.95\linewidth]{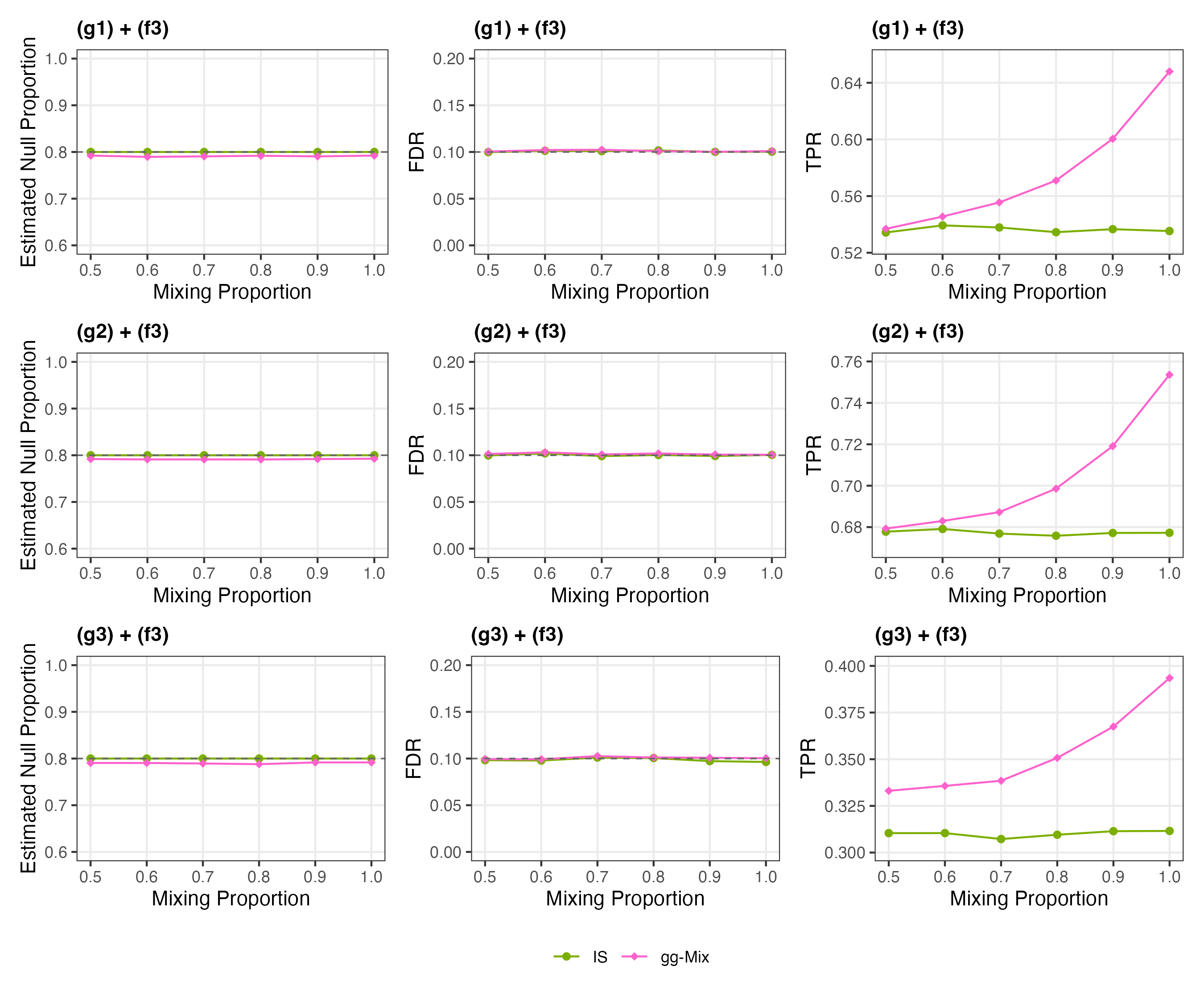}
    \caption{Effect of Structure Information of $ f $: Each row, from top to bottom, shows the results for priors (g1) to (g3) with a fixed $ f = (f3) $. In all panels, the x-axis represents the mixing proportion $ \pi_1^\prime $, while the y-axis differs depending on the column. In the left column, the y-axis represents the estimated null proportion, the middle columns display the FDR, and the right column shows the TPR. The horizontal dashed lines in the first and third columns indicate the true null proportion of 0.8 and the target FDR level of 0.1, respectively.}
    
    \label{fig:7}
\end{figure}

We compare the performance of gg-Mix with that of IS. For IS, we employ the oracle Storey-BH procedure \cite{storey2002direct} for multiplicity correction instead of the standard BH procedure. Specifically, we use the true value of the null proportion to adjust the conditional $p$-values. This approach enables us to assess the pure power gain achieved by explicitly modeling the non-null prior for $\mu_i$. The results are summarized in Figure~\ref{fig:7}, from which we note the following observations.

Both gg-Mix and IS control the FDR at the target level, but gg-Mix consistently shows higher power than IS across all simulation settings. As the mixing proportion $\pi_1'$ approaches 1, which indicates that the structure of the non-null prior $f$ becomes more informative, the gap in power between the two methods becomes more pronounced. 
Since the oracle Storey-BH procedure is used for multiplicity correction, the observed power loss of IS arises entirely because it does not account for the structural information of the non-null distribution $f$ in its testing procedure. This finding underscores the importance of incorporating the structural information of $f$ to enhance power.

\section{Real Data Example}\label{sec4}
In this section, we consider three real data examples previously analyzed in \cite{ignatiadis2025empirical}, and apply IS, ash-TwoStep, MixTwice, and gg-Mix to these datasets. We use the summary statistics provided by the authors of \cite{ignatiadis2025empirical} and compare the results across all methods. 
In addition to these real-data applications, we conduct a supplementary semi-synthetic analysis based on the gene-expression data of \cite{tian2003role}, following the approach of \cite{lu2019empirical}. This analysis allows us to evaluate FDR control and power with known ground truth while retaining important characteristics of real gene-expression data. See Section~\ref{appendix_g} of the Supplementary Material for details.

\subsection{Description of Real Datasets}

\subsubsection{Methylation Dataset (\cite{zhang2013genome})}
The DNA methylation dataset from \cite{zhang2013genome} comprises $\log_2$ ratios of methylated to unmethylated signal intensities measured at 439,918 CpG sites, using the Illumina HumanMethylation450 array. The data include 10 samples collected from three individuals, spanning four immune cell types (which include naive T-cells and antigen-activated naive T-cells).
The primary objective of the analysis is to identify CpG sites exhibiting significant differences in methylation levels between two cell types: naive T-cells and antigen-activated naive T-cells.
For each CpG site, summary statistics $(X_i, S_i^2)$ are computed by fitting a linear regression model that adjusts for both subject and cell type effects, yielding $\nu = 10 - 1 - (3 - 1) - (4 - 1) = 4$ degrees of freedom for the variance estimate. Since one hypothesis is assigned to each CpG site, the total number of hypotheses is $m = 439,918$.

\subsubsection{Gene Expression Dataset (\cite{dietrich2018drug})}
The gene expression dataset from \cite{dietrich2018drug} comprises measurements from 12 patients diagnosed with chronic lymphocytic leukemia (CLL). For each patient, paired samples were collected before and after a 12-hour incubation with Ibrutinib, a targeted therapy for CLL. Gene expression levels were profiled for 48,100 genes across these paired samples. 
The objective is to identify genes whose expression levels are significantly altered by Ibrutinib treatment. For each gene, $X_i$ denotes the average paired difference in expression levels across all 12 patients, and $S_i^2$ represents the estimated sampling variance. This yields $\nu = 12-1=11$ degrees of freedom for the variance estimate. Since one hypothesis is assigned to each gene, the total number of hypotheses is $m = 48,100$.

\subsubsection{Proteomics Dataset (\cite{terkelsen2021high})}
The proteomics dataset from \cite{terkelsen2021high} was generated using liquid chromatography-tandem mass spectrometry (LC-MS/MS) on 34 tumor interstitial fluid samples obtained from breast cancer patients, spanning three molecular subtypes: luminal, Her2, and triple-negative. The samples were processed in four experimental batches. For each of the 6,763 proteins, the outcome variable is the $\log_2$-transformed intensity, measured through LC-MS/MS against a pooled internal standard.
The primary objective of the analysis is to identify proteins that exhibit significant differences in expression between the luminal and Her2 subtypes. For each protein, the summary statistics $(X_i, S_i^2)$ are obtained by fitting a linear regression model that adjusts for breast cancer subtype and experimental batch, yielding $\nu = 34 - 1 - (3 - 1) - (4 - 1) = 28$ degrees of freedom for the variance estimate.
Since one hypothesis is assigned to each protein, the total number of hypotheses is $m = 6,763$.

\subsection{Results}
\begin{figure}[htb!]
    \centering
    \includegraphics[width=\linewidth]{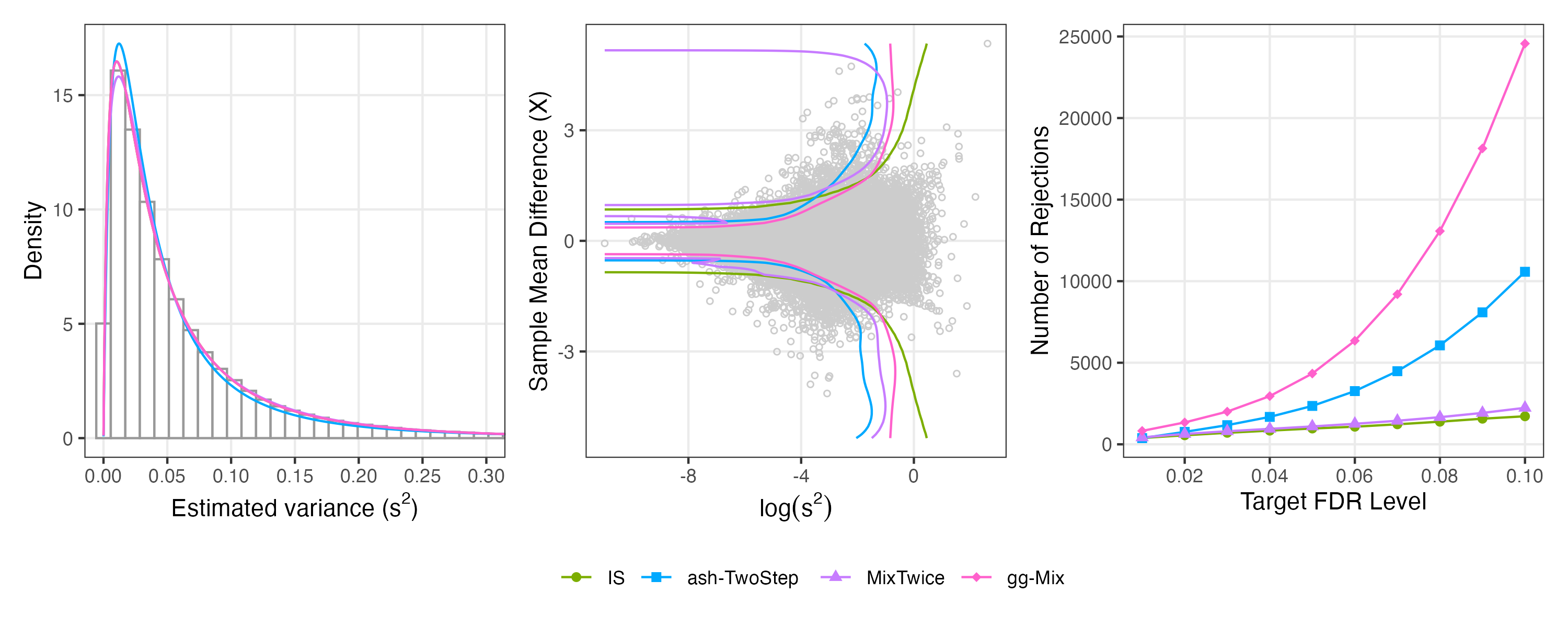}
    \caption{Analysis results on the Methylation dataset: The first panel displays the estimated marginal density of $S_i^2$ overlaid with its histogram. The second panel shows the rejection boundaries at the target FDR level of $0.05$, overlaid on the scatter plot of $(X_i, \log{S_i^2})$. Each gray circle represents a hypothesis; those with positive sample mean differences exceeding the upper rejection boundary, or negative ones falling below the lower boundary, are considered rejected. The third panel summarizes the number of rejections across varying target FDR levels from 0.01 to 0.1.}
    
    \label{fig:8}
\end{figure}

\begin{figure}[htb!]
    \centering
    \includegraphics[width=\linewidth]{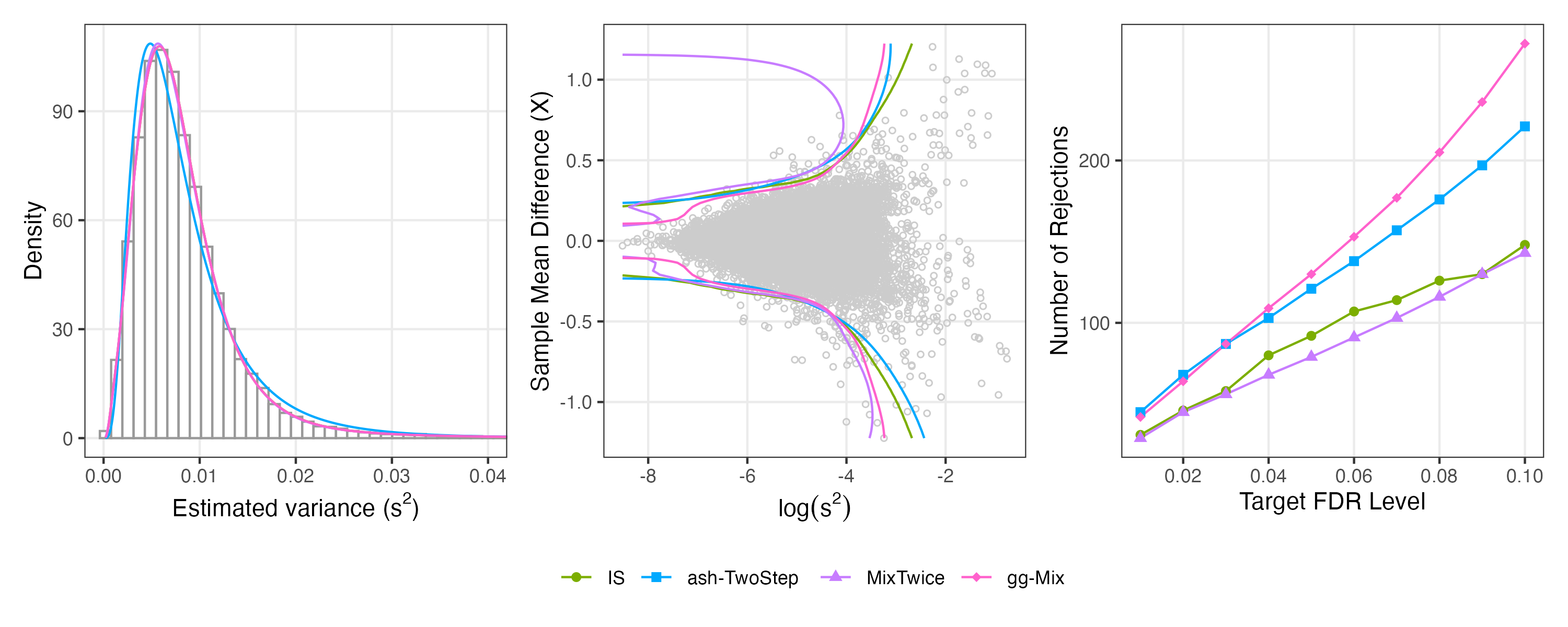}
    \caption{Analysis results on the Gene Expression dataset: The first panel displays the estimated marginal density of $S_i^2$ overlaid with its histogram. The second panel shows the rejection boundaries at the target FDR level of $0.05$, overlaid on the scatter plot of $(X_i, \log{S_i^2})$. Each gray circle represents a hypothesis; those with positive sample mean differences exceeding the upper rejection boundary, or negative ones falling below the lower boundary, are considered rejected. The third panel summarizes the number of rejections across varying target FDR levels from 0.01 to 0.1.}
    
    \label{fig:9}
\end{figure}

\begin{figure}[htb!]
    \centering
    \includegraphics[width=\linewidth]{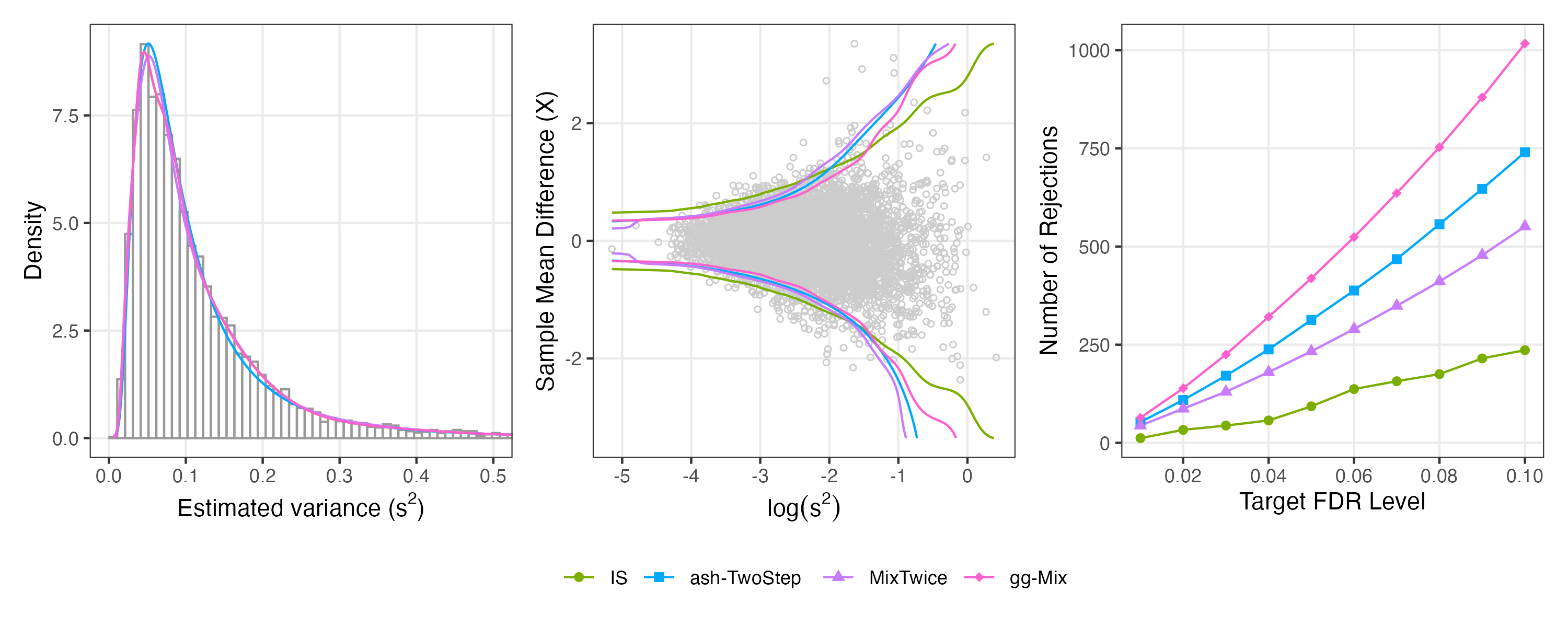}
    \caption{Analysis results on the Proteomics dataset: The first panel displays the estimated marginal density of $S_i^2$ overlaid with its histogram. The second panel shows the rejection boundaries at the target FDR level of $0.05$, overlaid on the scatter plot of $(X_i, \log{S_i^2})$. Each gray circle represents a hypothesis; those with positive sample mean differences exceeding the upper rejection boundary, or negative ones falling below the lower boundary, are considered rejected. The third panel summarizes the number of rejections across varying target FDR levels from 0.01 to 0.1.}
    
    \label{fig:10}
\end{figure}

The results for all three datasets are summarized in Figures~\ref{fig:8}--\ref{fig:10}. 

In each figure, the first panel overlays the histogram of $S_i^2$ (gray bars) with the estimated marginal density curves from each method.
Nonparametric approaches (IS, MixTwice, and gg-Mix), which do not impose a specific parametric form on the distribution of $\sigma_i^2$, show good alignment with the histogram.
In contrast, the parametric approach (ash-TwoStep) assumes a scaled inverse-chi-squared prior for $\sigma_i^2$, thereby limiting its flexibility. As a result, ash-TwoStep exhibits slight mismatches in its estimated density: it overestimates the density near the mode in the Methylation and Proteomics datasets, and underestimates it at certain points—around $S_i^2 = 0.1$ in the Methylation dataset and $S_i^2 = 0.18$ in the Proteomics dataset. In the Gene Expression dataset, it produces a slightly right-skewed estimate that does not align well with the histogram.

The second panel presents the rejection boundaries at the target FDR level of $0.05$, overlaid on the scatter plot of $(X_i, \log{S_i^2})$. In the Methylation dataset, IS, ash-TwoStep, MixTwice, and gg-Mix reject 969, 2{,}352, 1{,}170, and 4{,}337 hypotheses, respectively; in the Gene Expression dataset, they reject 92, 121, 79, and 130 hypotheses; and in the Proteomics dataset, 93, 313, 233, and 419 hypotheses, respectively. gg-Mix yields more rejections than IS, ash-TwoStep, and MixTwice across all three datasets. The rejection boundaries of gg-Mix differ substantially from those of the other methods in all three datasets, reflecting distinct underlying selection rules.

The third panel displays the number of rejections across target FDR levels ranging from $0.01$ to $0.1$. Across all datasets, gg-Mix yields the highest number of rejections at nearly all target FDR levels. At the few levels where ash-TwoStep yields slightly more rejections than gg-Mix, the numbers of rejections from ash-TwoStep and gg-Mix are very similar. IS and MixTwice identify fewer discoveries throughout the target FDR range.

\section{Conclusion}\label{sec5}
In this paper, we address the normal mean inference problem and propose a new empirical Bayes method, called gg-Mix. Our approach is distinguished from existing methods by its minimal prior assumptions for $\mu_i$ and $\sigma_i^2$: specifically, we remove commonly imposed distributional constraints, such as the scaled inverse-chi-squared assumption for $\sigma_i^2$ and the unimodality assumption for the non-null density of $\mu_i$. Through extensive numerical studies, we compare the performance of gg-Mix to that of existing methods. Our results show that gg-Mix consistently achieves the highest power while controlling the FDR at the target level across nearly all simulation settings considered. In particular, the validity of existing methods developed under the empirical Bayes framework is highly sensitive to the correctness of their modeling assumptions for $\mu_i$ and $\sigma_i^2$. Thus, when there is little or no information about the underlying prior distribution, gg-Mix is a safe and powerful choice.

We conclude the paper by discussing several limitations of our proposed method and outlining directions for future research.
First, our methodology is developed under the assumption that $\mu_i$ and $\sigma_i^2$ are independent. In practice, however, this assumption may be violated in various ways \citep{weinstein2018group, chen2026empirical}. Moreover, the hierarchical model specified in \eqref{eqn:assum_prior_1}--\eqref{eqn:like_2} assumes mutual independence among the observed pairs of summary statistics, which may not hold in real-world applications \citep{benjamini2001control, sun2009large, fan2017estimation, heller2021optimal, du2023false}. One promising direction for future research is therefore to extend the gg-Mix framework to accommodate potential dependence structures, both between $\mu_i$ and $\sigma_i^2$ and among the summary statistics themselves. Developing such extensions would enhance the practical applicability of gg-Mix in a broader range of real-world settings.
Second, our proposed method involves a tuning parameter $\lambda$, which controls the degree of conservativeness in estimating the null proportion. In this paper, we set $\lambda=10$ following the recommendation of \cite{stephens2017false}; however, this choice is not necessarily optimal. Developing a data-adaptive strategy for selecting $\lambda$ remains an important direction for future research.
Finally, establishing theoretical guarantees for gg-Mix remains an important issue. In a related empirical partially Bayes setting, \citet{ignatiadis2025empirical} established consistency of the NPMLE of $G$ and asymptotic validity of the resulting multiple testing procedure. However, gg-Mix introduces several additional theoretical challenges. First, as an empirical Bayes procedure, gg-Mix requires consistent estimation of the non-null density $f$ in addition to $G$. Second, because gg-Mix incorporates regularization to guard against underestimation of $\pi_0$, the effect of the penalty parameter $\lambda$ on the estimation of both $f$ and $\pi_0$ needs to be accounted for. Third, gg-Mix estimates $G$ on a prespecified finite grid rather than using the unrestricted NPMLE, introducing an additional approximation error that needs to be characterized. Finally, establishing FDR control requires understanding how the estimation and approximation errors in $G$, $f$, and $\pi_0$ propagate to the estimated local false discovery rates and the resulting data-adaptive threshold. Addressing all of these issues requires substantial additional theoretical development. We therefore leave these for future research.

\section*{Conflict of Interest}
The authors declare no conflicts of interest.

\section*{Funding statement}
This work was supported by National Research Foundation of Korea (NRF) (No. RS-2025-00520739).

\section*{Data and Code Availability}
The data analyzed in this study are available from the repository of \cite{ignatiadis2025empirical} at \href{https://github.com/nignatiadis/empirical-partially-bayes-paper}{GitHub}. The code for gg-Mix and the real data analyses is publicly available \href{https://kwangokseo.vercel.app/gg_Mix.zip}{here}.

\bibliographystyle{chicago}
\bibliography{ref}

\clearpage

\begin{center}
    \LARGE
    Supplementary Material for\\[0.3em]
    ``Empirical Bayes Method for Large-Scale Multiple Testing with Heteroscedastic Errors''
\end{center}

\appendix

\numberwithin{equation}{section}
\numberwithin{figure}{section}

\section{Density Functions Related to the Proposed Method}\label{appendix_a}

In this section, we summarize the closed form of density functions related to our proposed method. Throughout this section, we assume that the assumptions outlined in \eqref{eqn:assum_prior_1}--\eqref{eqn:like_2} hold.

\subsection{An Approximation of the Marginal Density of $S_i^2$}
\begin{align*}
    p(S_i^2 ; \boldsymbol{\Delta})
    &= \int p(S_i^2 \mid \sigma_i^2) ~ d\tilde{G}(\sigma_i^2; \boldsymbol{\Delta}) \\
    &= \int \text{Gamma}\left(S_i^2; \frac{\nu}{2}, \frac{\nu}{2\sigma_i^2}\right) \cdot \sum_{l=1}^L\Delta_l \delta_{\kappa_l}(d\sigma_i^2) \\
    &= \sum_{l=1}^L\Delta_l \cdot \text{Gamma}\left(S_i^2; \frac{\nu}{2}, \frac{\nu}{2\kappa_l}\right),
\end{align*}
where $\text{Gamma}(\cdot; \alpha, \beta)$ denotes the gamma density function with shape parameter $\alpha$ and rate parameter $\beta$.

\subsection{An Approximation of the Posterior Density of $\sigma_i^2$}
\begin{align*}
    p(\sigma_i^2 \mid S_i^2; \boldsymbol{\Delta})
    &= \frac{p(S_i^2 \mid \sigma_i^2) ~ d\tilde{G}(\sigma_i^2; \boldsymbol{\Delta})}
    {p(S_i^2; \boldsymbol{\Delta})}\\
    &= \frac{\text{Gamma}(S_i^2; \frac{\nu}{2}, \frac{\nu}{2\sigma_i^2}) \cdot \sum\limits_{l=1}^L\Delta_l \delta_{\kappa_l}(d\sigma_i^2)}{\sum\limits_{l=1}^L\Delta_l \cdot 
    \text{Gamma} (S_i^2; \frac{\nu}{2}, \frac{\nu}{2\kappa_l})}.
\end{align*}

\subsection{An Approximation of the Conditional Density of $X_i$ Given $\mu_i$ and $S_i^2$}
\begin{align*}
    p(X_i \mid \mu_i, S_i^2; \boldsymbol{\Delta})
    &= \int p(X_i \mid \mu_i, \sigma_i^2)
    \cdot p(\sigma_i^2 \mid S_i^2; \boldsymbol{\Delta}) ~d\sigma_i^2\\
    &= \int \frac{1}{\sqrt{2\pi \sigma_i^2}}e^{-\frac{1}{2\sigma_i^2}\left(X_i-\mu_i\right)^2}
    \cdot \frac{\text{Gamma}(S_i^2; \frac{\nu}{2}, \frac{\nu}{2\sigma_i^2}) \cdot \sum\limits_{l=1}^L\Delta_l \delta_{\kappa_l}(d\sigma_i^2)}{\sum\limits_{l=1}^L\Delta_l \cdot \text{Gamma} (S_i^2; \frac{\nu}{2}, \frac{\nu}{2\kappa_l})}\\
    &= \sum_{l=1}^L 
    \frac{\Delta_l \cdot 
    \text{Gamma}\left(S_i^2; \frac{\nu}{2}, \frac{\nu}{2\kappa_l}\right)}
    {\sum\limits_{l=1}^L\Delta_l \cdot 
    \text{Gamma}\left(S_i^2; \frac{\nu}{2}, \frac{\nu}{2\kappa_l}\right)}
    \cdot \frac{1}{\sqrt{2\pi \kappa_l}}e^{-\frac{1}{2\kappa_l}\left(X_i-\mu_i\right)^2}\\
    &= \sum_{l=1}^L \pi_{i,l} \cdot N\left(X_i; \mu_i, \kappa_l\right),
\end{align*}
where 
\begin{align*}
    \pi_{i,l} 
    \coloneqq \pi_{i,l}(S_i^2; \boldsymbol{\Delta})
    =\frac{\Delta_l \cdot 
    \text{Gamma}\left(S_i^2; \frac{\nu}{2}, \frac{\nu}{2\kappa_l}\right)}
    {\sum\limits_{l=1}^L\Delta_l \cdot 
    \text{Gamma}\left(S_i^2; \frac{\nu}{2}, \frac{\nu}{2\kappa_l}\right)}.
\end{align*}

\subsection{An Approximation of the Conditional Density of $X_i$ Given $S_i^2$}
\begin{align*}
    &p(X_i \mid S_i^2; \boldsymbol{\Delta}, \boldsymbol{\pi})\\
    &= \int p(X_i \mid \mu_i, S_i^2; \boldsymbol{\Delta})
    \cdot p(\mu_i; \boldsymbol{\pi})~d\mu_i\\
    &= \int  p(X_i \mid \mu_i, S_i^2; \boldsymbol{\Delta}) \cdot 
    (1-\pi) \cdot \delta_0(\mu_i)~d\mu_i + \int p(X_i \mid \mu_i, S_i^2; \boldsymbol{\Delta}) \cdot \pi \cdot \tilde{f}(\mu_i)~d\mu_i\\
    &= (1-\pi) \cdot \left(\sum_{l=1}^L\pi_{i,l}
    \cdot \frac{1}{\sqrt{2\pi \kappa_l}}e^{-\frac{1}{2\kappa_l}X_i^2} \right)
    + \pi \cdot \left[ \sum_{k=1}^K\pi_k^\prime \cdot 
    \left\{\sum_{l=1}^L\pi_{i,l} \cdot
    \int \frac{1}{\sqrt{2\pi \kappa_l}}e^{-\frac{1}{2\kappa_l}\left(X_i-\mu_i\right)^2}
    \cdot f_k(\mu_i) ~d\mu_i\right\} \right]\\
    &= \pi_0 \cdot \left(\sum_{l=1}^L\pi_{i,l}
    \cdot \frac{1}{\sqrt{2\pi \kappa_l}}e^{-\frac{1}{2\kappa_l}X_i^2} \right)
    + \sum_{k=1}^K\pi_k \cdot 
    \left\{\sum_{l=1}^L\pi_{i,l} \cdot
    \int \frac{1}{\sqrt{2\pi \kappa_l}}e^{-\frac{1}{2\kappa_l}\left(X_i-\mu_i\right)^2}
    \cdot f_k(\mu_i) ~d\mu_i\right\},
\end{align*}
where $\tilde{f}(\mu_i) = \sum_{k = 1}^K \pi_k^\prime \cdot f_k(\mu_i)$.
Depending on the form of $f_k$, we have 
\begin{align*}
\int \frac{1}{\sqrt{2\pi \kappa_l}}e^{-\frac{1}{2\kappa_l}\left(X_i-\mu_i\right)^2}
    \cdot f_k(\mu_i) ~d\mu_i = 
    \begin{cases} 
        N\left(X_i; \gamma_k, \zeta_k^2 + \kappa_l\right),
        & \text{if } f_k(\cdot) = N(\cdot ;\gamma_k, \zeta_k^2)\\\\
        \dfrac{\Phi\left(\frac{X_i-\alpha_k}{\sqrt{\kappa_l}}\right)
        -\Phi\left(\frac{X_i-\beta_k}{\sqrt{\kappa_l}}\right)}{\beta_k-\alpha_k}, 
        & \text{if } f_k(\cdot) = U(\cdot ;\alpha_k, \beta_k),
    \end{cases}
\end{align*}
where $ \Phi(\cdot) $ denotes the distribution function  of the standard normal distribution.

\section{Sensitivity of gg-Mix to the Number of Grid Points}\label{appendix_b}
In this section, we focus on the case where the Gaussian location $+$ Gaussian scale mixture is used as the working model for $f$ and examine the sensitivity of gg-Mix to the number of grid points. In implementing gg-Mix, two types of grids need to be specified: one for $G$, and the other for the Gaussian location mixture component of $f$ (the grid for the Gaussian scale mixture is selected automatically, following \citet{stephens2017false}).

\begin{figure}[!htb]
    \centering
    \includegraphics[width=\linewidth]{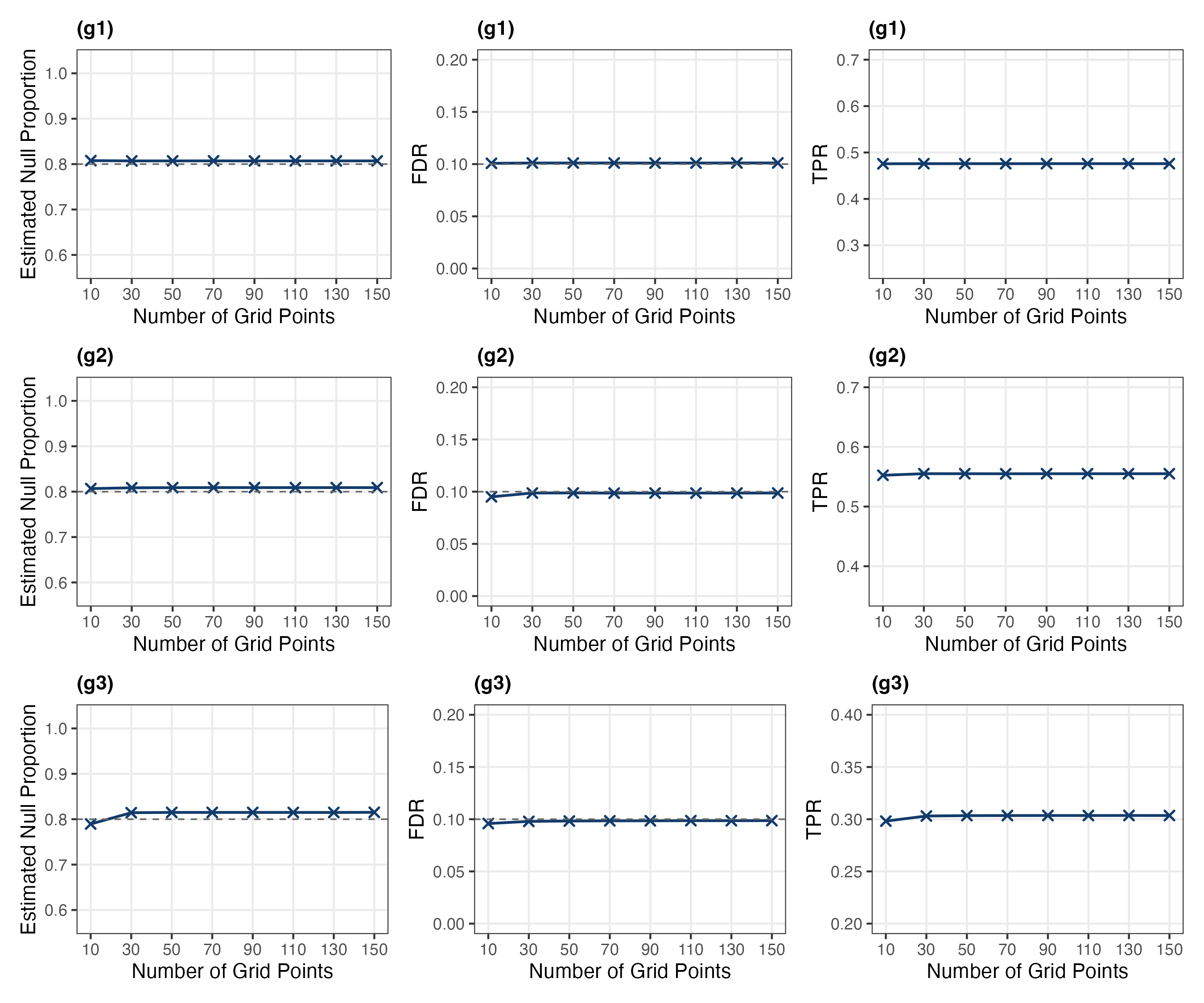}
    \caption{Sensitivity of the estimated null proportion, FDR, and TPR to the number of grid points for \(G\).}
    \label{fig:B1}
\end{figure}

First, we examine the sensitivity to the number of grid points for $G$ across different forms of the underlying distribution $G$. Fixing the number of grid points for the Gaussian location mixture at 50, the number of grid points for $G$ is varied from 10 to 150, with $(\mathrm{g1})$--$(\mathrm{g3})$ considered for $G$. The remaining setting consists of $m=5{,}000$ hypotheses with $\nu=10$, $\pi_0=0.8$, and a target FDR level of $0.1$, with $f$ fixed at $(\mathrm{f1})$. The estimated null proportion, FDR, and TPR are each averaged over 200 replications. The results are summarized in Figure~\ref{fig:B1}.

Across all three choices of $G$, the estimated null proportion, FDR, and TPR remain nearly unchanged as the number of grid points for $G$ varies. The estimated null proportion remains close to the true value, and the FDR is controlled near the target level. In particular, $L=50$ provides stable performance across all three choices of $G$.
    
\begin{figure}[!htb]
    \centering
    \includegraphics[width=\linewidth]{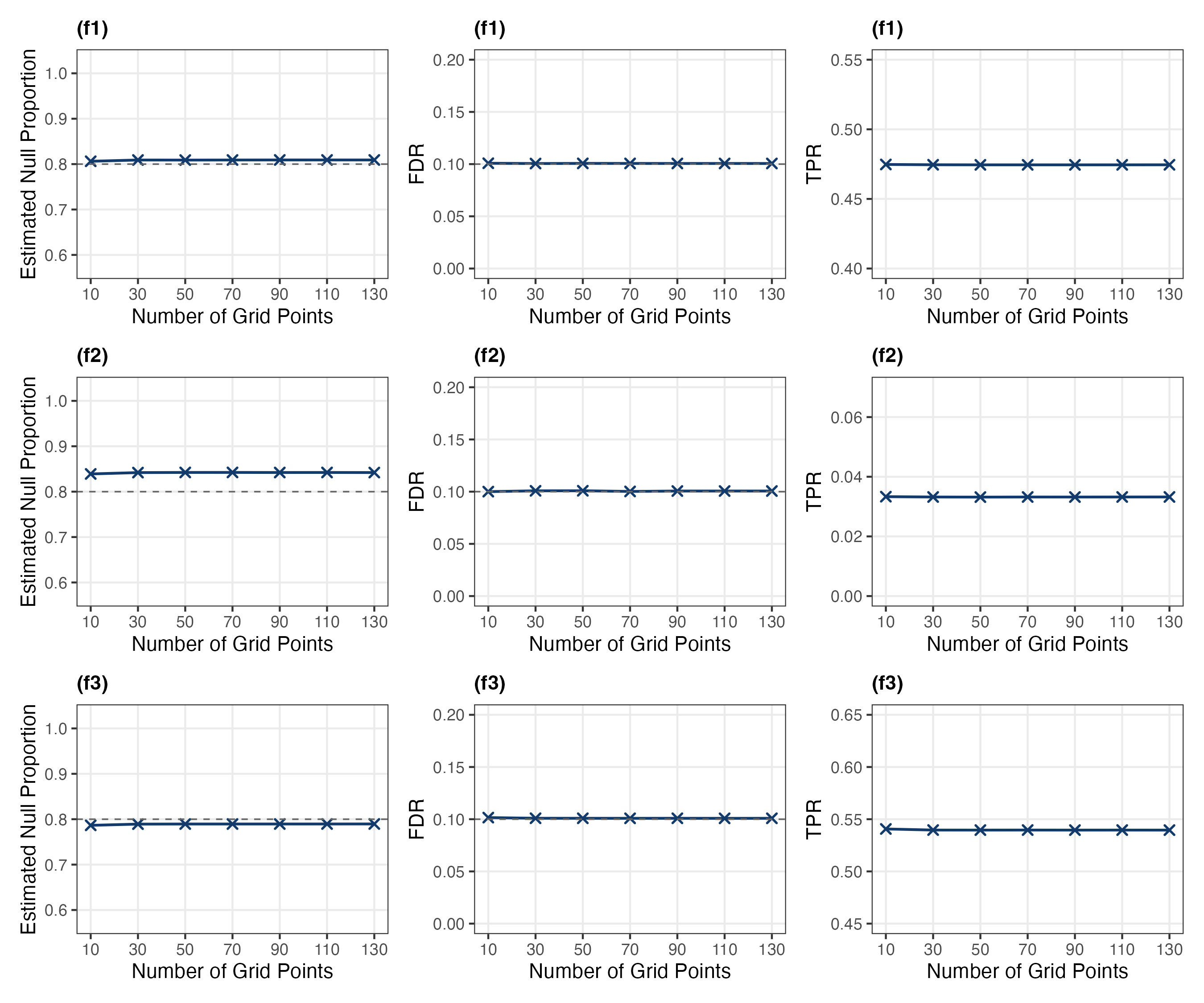}
    \caption{Sensitivity of the estimated null proportion, FDR, and TPR to the number of grid points for the Gaussian location mixture.}
    \label{fig:B2}
\end{figure}

Second, we examine the sensitivity to the number of grid points for the Gaussian location mixture across different forms of the non-null distribution $f$. Fixing the number of grid points for $G$ at 50, the number of grid points for the Gaussian location mixture is varied from 10 to 130, with $(\mathrm{f1})$--$(\mathrm{f3})$ considered for $f$ and $G$ fixed at $(\mathrm{g1})$. All other simulation settings and evaluation criteria remain the same as in the preceding analysis. The results are summarized in Figure~\ref{fig:B2}.

We observe a similar pattern. The estimated null proportion, the FDR, and the TPR are again almost identical across the grid sizes considered, confirming that gg-Mix is robust to this choice as well. In particular, $K_1=50$ provides stable performance across all three choices of $f$.

Taken together, these two studies indicate that gg-Mix is not particularly sensitive to the number of grid points. In the settings considered, 50 grid points are sufficient for both $G$ and the Gaussian location mixture component of $f$.

\section{Sensitivity of gg-Mix to the Choice of $\lambda$}\label{appendix_c}
In this section, we examine the sensitivity of gg-Mix to the choice of the penalty parameter $\lambda$, which controls the strength of regularization on the null proportion, and investigate its effect on null proportion estimation, FDR control, and power.

\begin{figure}[!htb]
    \centering
    \includegraphics[width=\linewidth]{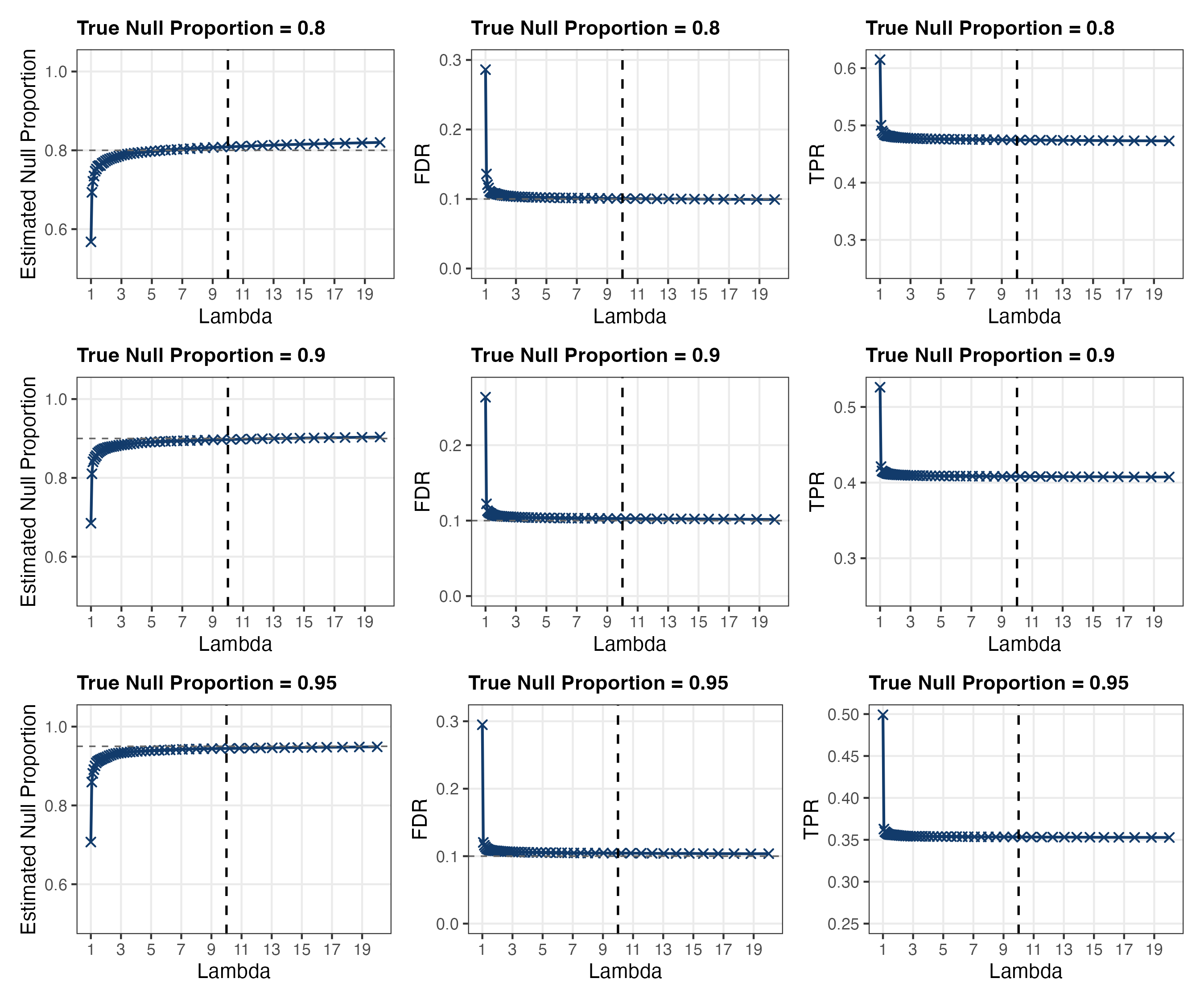}
    \caption{Sensitivity of the estimated null proportion, FDR, and TPR to the regularization parameter $\lambda$.}
    \label{fig:C1}
\end{figure}

To this end, we vary $\lambda$ over a grid on $[1,20]$ and consider $\pi_0\in\{0.8,0.9,0.95\}$, with $G=(\mathrm{g1})$ and $f=(\mathrm{f1})$. The remaining settings are $m=5{,}000$, $\nu=10$, and a target FDR level of $0.1$. The estimated null proportion, FDR, and TPR are each averaged over 200 replications. The results are summarized in Figure~\ref{fig:C1}.
    
For all $\lambda \geq 3$, the estimated null proportion, the FDR, and the TPR vary only slightly with $\lambda$. Although all three quantities change monotonically as $\lambda$ increases, the changes are small, and their values remain nearly constant over this range for every true null proportion considered. Furthermore, the estimated null proportion remains close to the true value and the FDR is controlled at the target level. In particular, our default choice of $\lambda = 10$ (indicated by the vertical dashed line) appears to be reasonable, at least in the settings considered here.

In contrast, when $\lambda < 3$, all three quantities depend noticeably on the choice of $\lambda$. In particular, when $\lambda = 1$, which corresponds to imposing no regularization on the null proportion, $\pi_0$ is substantially underestimated across all true null proportions, and the FDR is inflated as a consequence.
    
These results show that gg-Mix is not sensitive to the choice of $\lambda$, provided that $\lambda$ is not too small. Furthermore, the results clarify the role of the penalty term in the settings considered here: it guards against underestimation of the null proportion, thereby helping to maintain FDR control.

\section{Computational Time of gg-Mix}\label{appendix_d}
To assess the computational scalability of gg-Mix, we conduct a numerical study. We increase the number of hypotheses $m$ from $5{,}000$ to $150{,}000$ and record the average computing time over 100 replications, comparing gg-Mix with ash-TwoStep. 

\begin{figure}[!htb]
    \centering
    \includegraphics[width=0.6\linewidth]{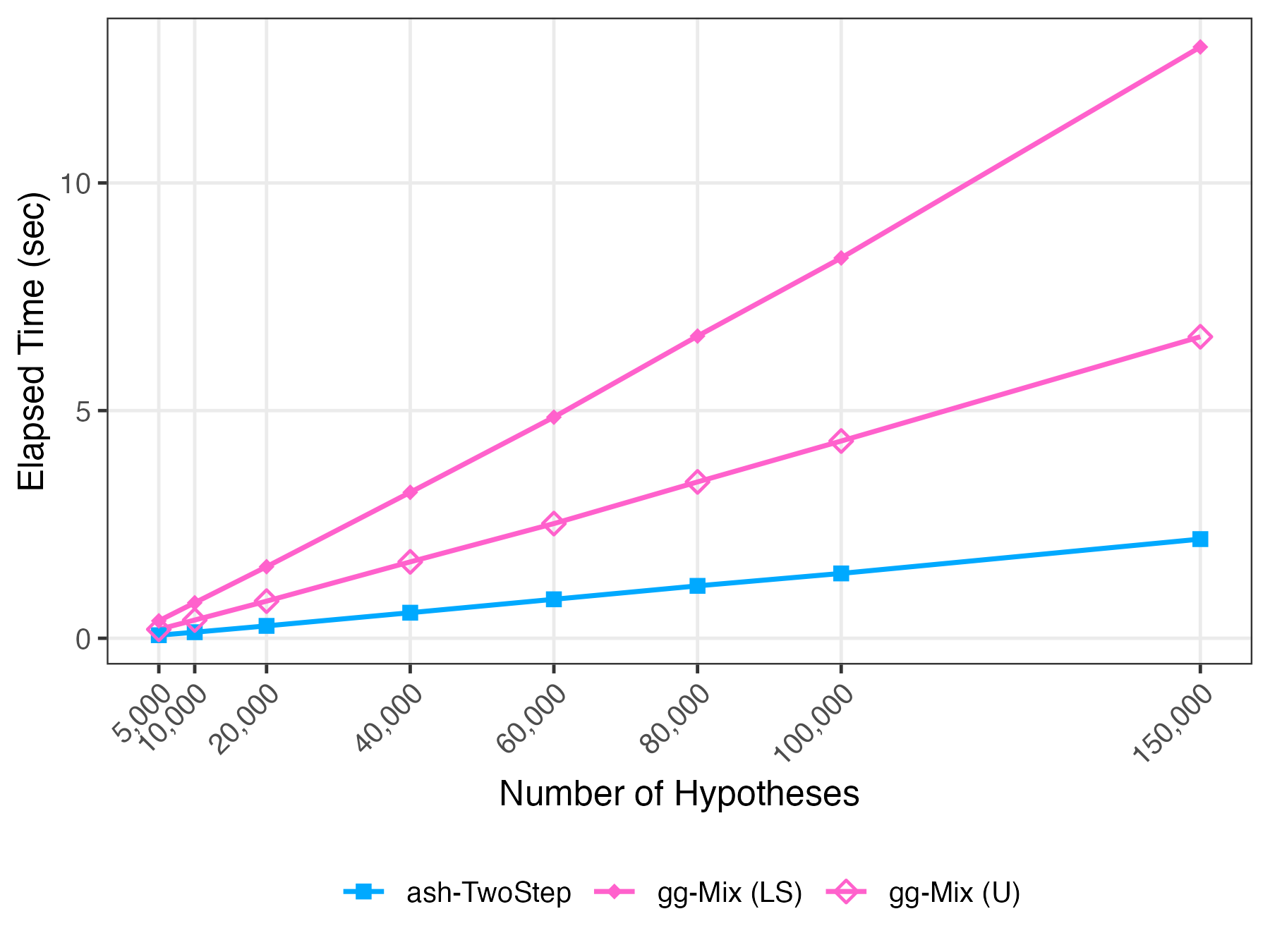}
    \caption{Average computation time as a function of the number of hypotheses.}
    \label{fig:D1}
\end{figure}

Since the computing time depends heavily on the working model for $f$, we run gg-Mix under two working models: the uniform mixture, denoted gg-Mix (U), and the Gaussian location $+$ Gaussian scale mixture, denoted gg-Mix (LS), which is our default recommendation. ash-TwoStep is run with the uniform mixture. The grid for the unimodal mixture component is selected automatically following the procedure proposed by \citet{stephens2017false}. For gg-Mix (LS), we use 50 grid points for the Gaussian location mixture component. For the remaining settings, we use $(\mathrm{g1})$ for $G$ and $(\mathrm{f1})$ for $f$, with $\pi_0=0.8$, and fix the number of grid points for $G$ at $L=50$ for both variants of gg-Mix. The results are summarized in Figure~\ref{fig:D1}.

Across all numbers of hypotheses, gg-Mix (LS) is the slowest and ash-TwoStep the fastest, with gg-Mix (U) in between. Since gg-Mix (U) and ash-TwoStep use the same uniform mixture to model $f$, the difference in computing time between them mainly reflects how the prior $G$ is estimated: ash-TwoStep relies on a parametric model for $G$, whereas gg-Mix estimates $G$ nonparametrically. The additional computational cost of gg-Mix (LS) relative to gg-Mix (U) arises primarily from the Gaussian location mixture component, which introduces additional mixture weights to be estimated. 

Nevertheless, the computing time grows approximately linearly with the number of hypotheses for all three procedures. Even with $150{,}000$ hypotheses, gg-Mix (LS) takes only about 13 seconds on average, suggesting that its additional computational cost is unlikely to be a practical concern. The favorable computational scaling of gg-Mix is partly due to the fact that the dimension of the optimization problem is determined by the grid sizes rather than by the number of hypotheses $m$. Thus, when the grid sizes are fixed, the number of parameters to be estimated does not increase with $m$, although the computational cost of evaluating the likelihood increases as $m$ grows.

\section{Complete Simulation Results Including Additional Methods}\label{appendix_e}
In this section, we present the complete simulation results, including additional methods omitted from the main manuscript for clarity. Alongside the four methods considered in the main manuscript, we include the Benjamini--Hochberg procedure \citep{benjamini1995controlling}, referred to as `BH', and the method of \cite{smyth2004linear}, referred to as `GS'. For BH, we compute $p$-values based on the null distribution $t_i=X_i/S_i\sim T_\nu$, where $T_\nu$ denotes the $t$-distribution with $\nu$ degrees of freedom, and apply the BH procedure to these $p$-values. For GS, we compute moderated $t$-statistics under a scaled inverse-chi-squared prior for $\sigma_i^2$ and apply the BH procedure to the resulting $p$-values. We also include two versions of the method proposed by \cite{stephens2017false}: one based on a normal approximation to the conditional distribution of $X_i\mid\mu_i,S_i^2$, referred to as `ash-n', and the other based on a $t$ approximation, referred to as `ash-t'.

\subsection{Effect of Null Proportion $\pi_0$}\label{appendix_e1}

\begin{figure}[htb!]
    \centering
    \includegraphics[width=\linewidth]{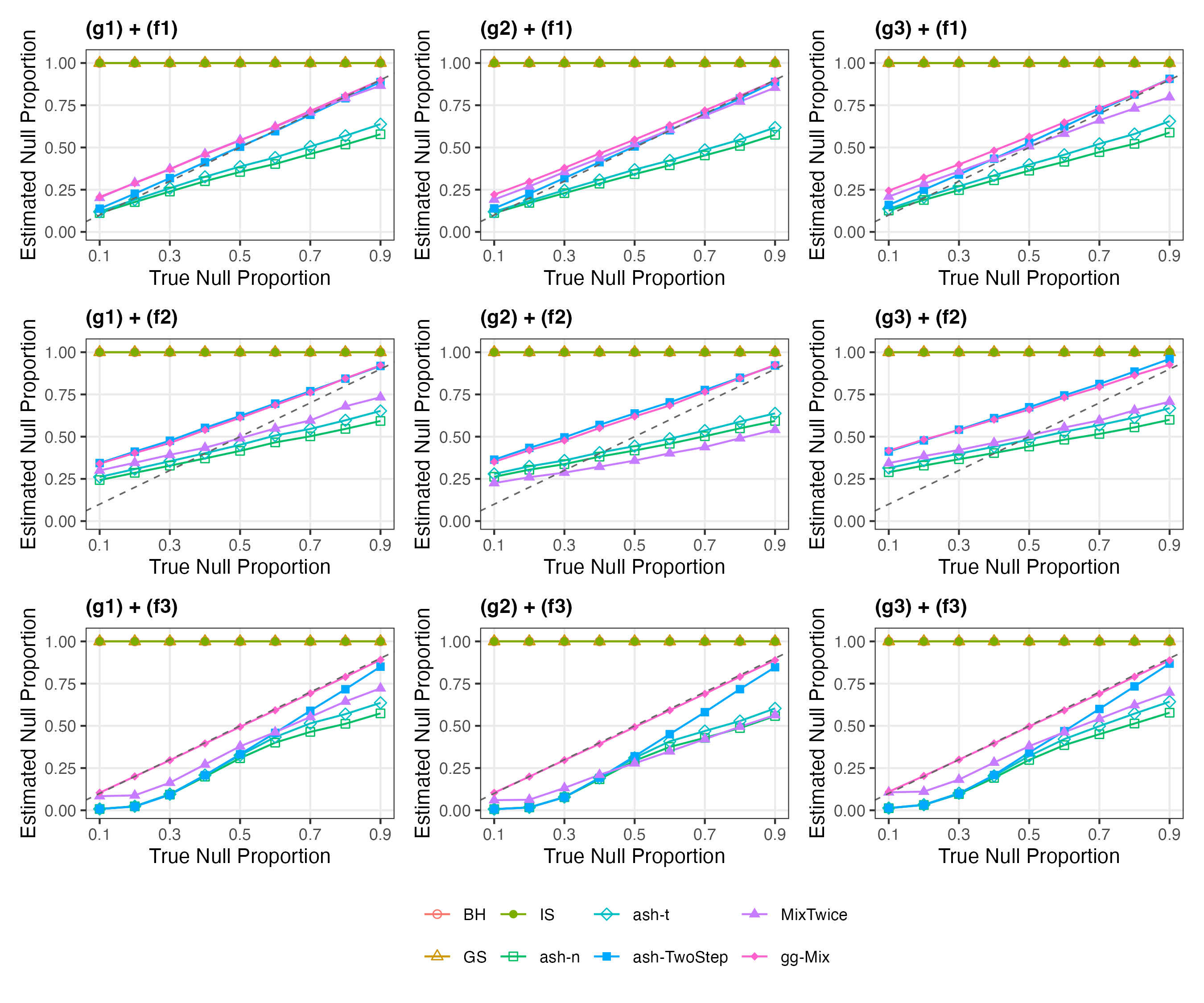}
    \caption{Estimated Null Proportion: Each column, from left to right, corresponds to the results for priors (g1) to (g3), while each row, from top to bottom, corresponds to priors (f1) to (f3). In each panel, the x-axis represents the true null proportion, and the y-axis shows the estimated null proportion. The black dashed lines indicate the 45-degree line.}
    \label{fig:E1}
\end{figure}

\begin{figure}[htb!]
    \centering
    \includegraphics[width=\linewidth]{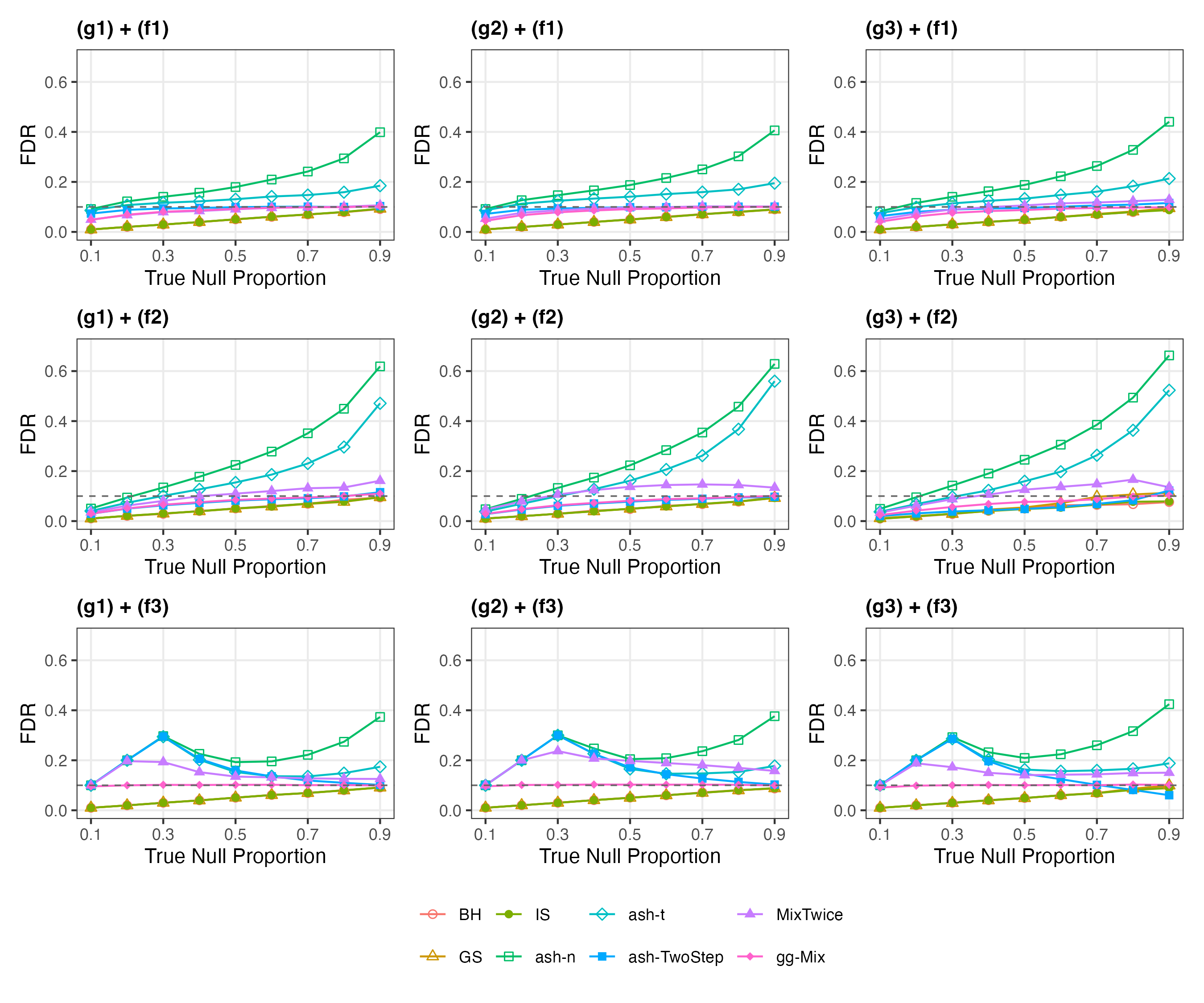}
    \caption{
    (Empirical) False Discovery Rate: Each column, from left to right, corresponds to the results for priors (g1) to (g3), while each row, from top to bottom, corresponds to priors (f1) to (f3). In each panel, the x-axis represents the null proportion, and the y-axis shows the FDR. The horizontal dashed lines represent the target FDR level of 0.1.}
    \label{fig:E2}
\end{figure}

\begin{figure}[htb!]
    \centering
    \includegraphics[width=\linewidth]{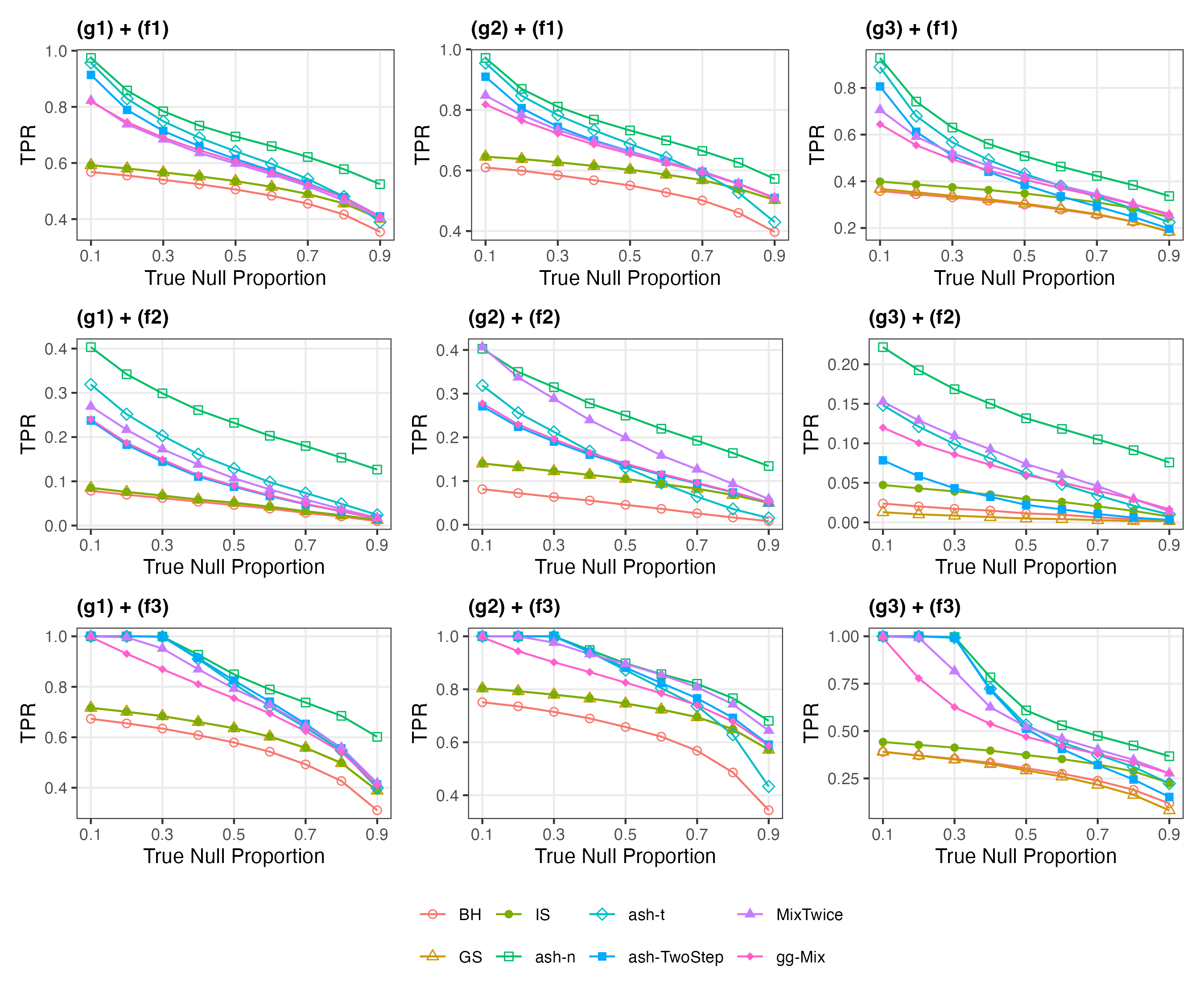}
    \caption{(Empirical) True Positive Rate: Each column, from left to right, corresponds to the results for priors (g1) to (g3), while each row, from top to bottom, corresponds to priors (f1) to (f3). In each panel, the x-axis represents the null proportion, and the y-axis shows the TPR.}
    \label{fig:E3}
\end{figure}

Recall that we fix $\nu = 10$ and $\pi_1' = 0.5$, and consider all possible combinations of the three types of $G$, the three types of $f$, and the nine levels of $\pi_0$, which yields a total of $3 \times 3 \times 9 = 81$ simulation settings. The results are summarized in Figures~\ref{fig:E1}--\ref{fig:E3}, from which we make the following observations.

First, BH and GS effectively use a null proportion of $1$, as does IS, since all three rely on the BH procedure for multiplicity correction. 

Second, BH, GS, and IS control the FDR across the simulation settings considered, but their power exhibits different patterns depending on the underlying setting. When the scaled inverse-chi-squared assumption is satisfied, GS and IS exhibit nearly identical power, and both outperform BH. This improvement over BH arises because GS and IS account for heterogeneity in $\sigma_i^2$, whereas BH does not. However, when the parametric assumption of GS is violated, its power can deteriorate substantially and may even fall below that of BH. For example, under $(g3)+(f2)$, BH achieves higher power than GS. In contrast, IS consistently outperforms both methods. This advantage arises from its nonparametric modeling of the distribution of $\sigma_i^2$, which allows IS to accommodate the underlying heterogeneity without relying on the scaled inverse-chi-squared assumption.

Third, ash-n substantially underestimates the null proportion and fails to control the FDR in most simulation settings. ash-t generally yields estimates of the null proportion and FDR values closer to the true and target levels, respectively, than ash-n. Nevertheless, ash-t still fails to control the FDR in most settings because the $t$ approximation does not fully account for the discrepancy between $S_i^2$ and the true variance $\sigma_i^2$. This discrepancy can be substantial in the setting considered here, where $\nu=10$.

\subsection{Effect of Degrees of Freedom $\nu$}
\begin{figure}[htb!]
    \centering
    \includegraphics[width=\linewidth]{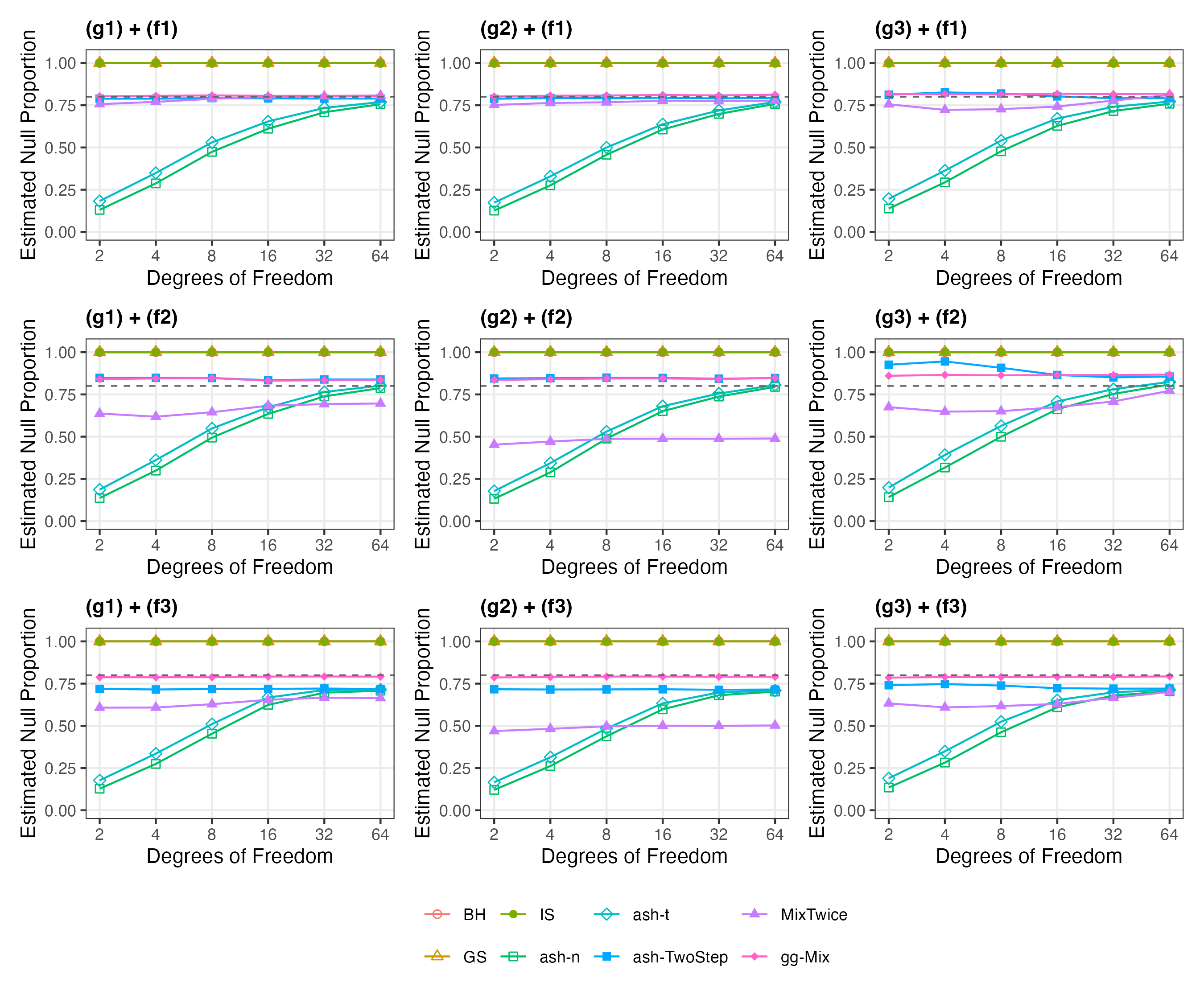}
    \caption{Estimated Null Proportion: Each column, from left to right, corresponds to the results for priors (g1) to (g3), while each row, from top to bottom, corresponds to priors (f1) to (f3). In each panel, the x-axis represents the degrees of freedom, and the y-axis shows the estimated null proportion. The black dashed lines indicate the true null proportion of 0.8.}
    \label{fig:E4}  
\end{figure}

\begin{figure}[htb!]
    \centering
    \includegraphics[width=\linewidth]{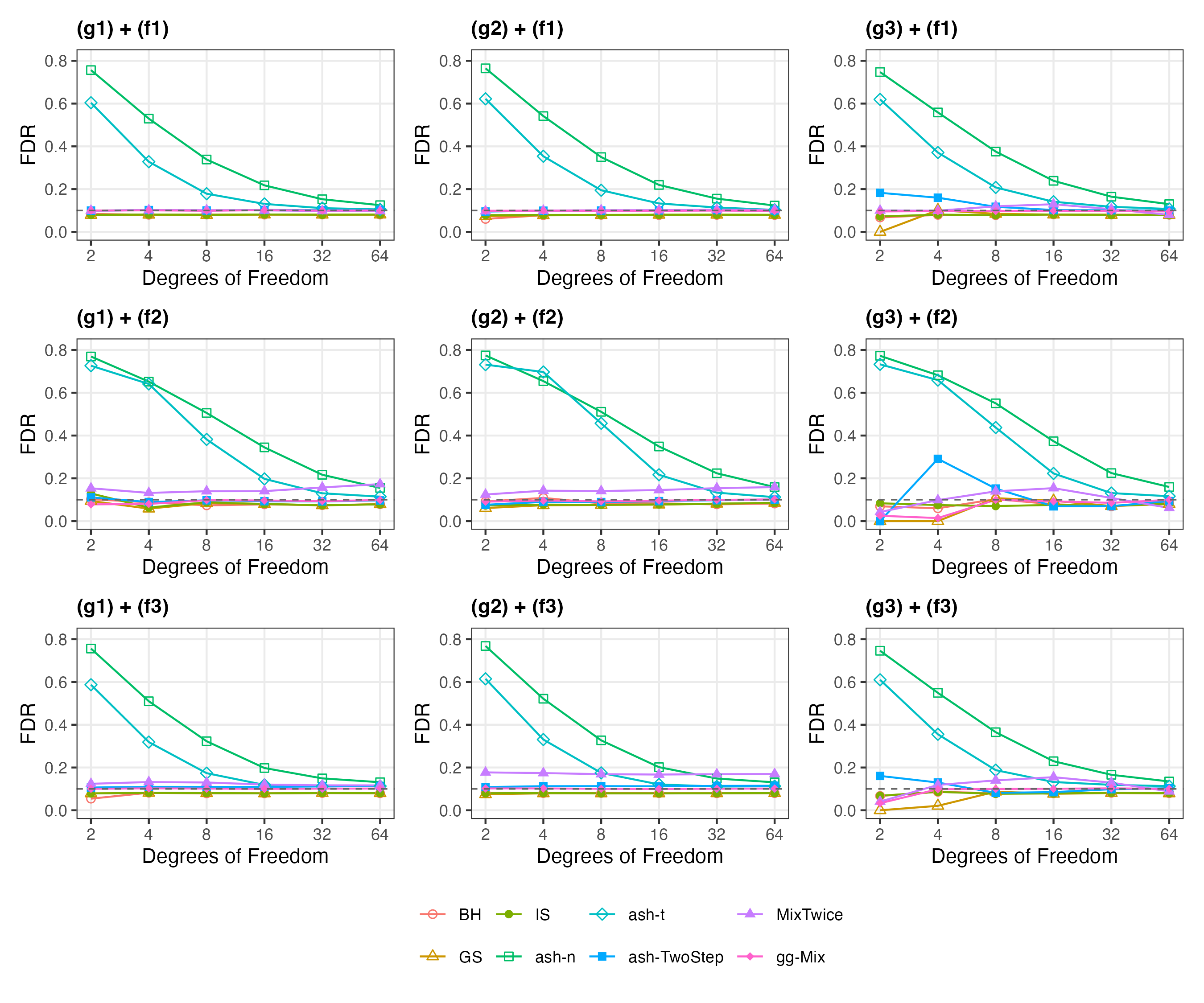}
    \caption{(Empirical) False Discovery Rate: Each column, from left to right, corresponds to the results for priors (g1) to (g3), while each row, from top to bottom, corresponds to priors (f1) to (f3). In each panel, the x-axis represents the degrees of freedom, and the y-axis shows the FDR. The horizontal dashed lines represent the target FDR level of 0.1.}
    \label{fig:E5}
\end{figure}

\begin{figure}[htb!]
    \centering
    \includegraphics[width=\linewidth]{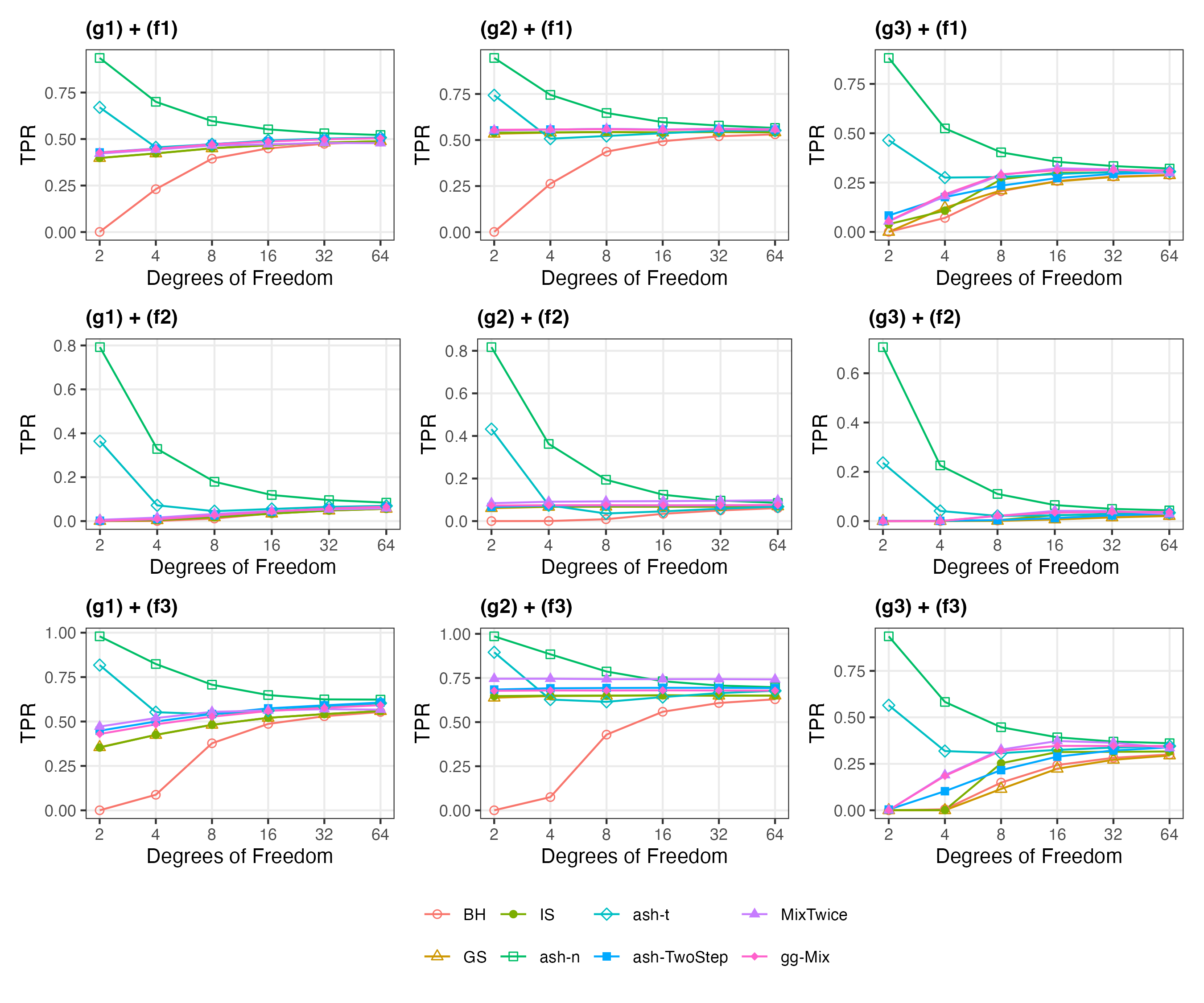}
    \caption{(Empirical) True Positive Rate: Each column, from left to right, corresponds to the results for priors (g1) to (g3), while each row, from top to bottom, corresponds to priors (f1) to (f3). In each panel, the x-axis represents the degrees of freedom, and the y-axis shows the TPR.}
    \label{fig:E6}
\end{figure}

Recall that we fix $\pi_0 = 0.8$ and $\pi_1' = 0.2$, and consider all possible combinations of the three types of $G$, the three types of $f$, and six levels of $\nu$, which yields a total of $3 \times 3 \times 6 = 54$ simulation settings. The results are summarized in Figures~\ref{fig:E4}--\ref{fig:E6}.  

We observe a pattern similar to that in Section~\ref{appendix_e1}. BH and GS control the FDR across all simulation settings, although they are generally less powerful than IS. In contrast, ash-n and ash-t fail to control the FDR in most settings when the degrees of freedom are small. However, as $\nu$ increases, both methods estimate the null proportion more accurately, and their FDRs approach the target level. In particular, when $\nu=64$, both methods provide relatively accurate estimates of the null proportion and maintain the FDR close to the target level across all simulation settings. This improvement occurs because, for large $\nu$, $S_i^2$ provides a more accurate estimate of $\sigma_i^2$.

\section{Additional Simulation Results}\label{appendix_f}
\subsection{Violation of Independence between $\mu_i$ and $\sigma_i^2$}
Many existing methods for the normal mean inference problem are developed under the assumption that $\mu_i$ and $\sigma_i^2$ are independent \citep{stephens2017false, lu2019empirical, zheng2021mixtwice, fu2022heteroscedasticity}. However, there are settings in which the two quantities are systematically related; for instance, genes with larger variances may also tend to exhibit larger effect sizes. 

When gg-Mix is applied to data in which $\mu_i$ and $\sigma_i^2$ are dependent, the working prior $\pi_0\delta_0 + (1-\pi_0)\tilde{f}$ is fitted as if it were common to all $i$, whereas the true non-null density of the effect size depends on the variance, that is, $f(\cdot \mid \sigma_i^2)$ varies with $\sigma_i^2$. The resulting misspecification can distort the estimated local false discovery rates, thereby affecting FDR control: the procedure may become either anti-conservative or overly conservative, depending on the direction of the misspecification.

In many applications, however, the non-null proportion $\pi$ is small. Since the non-null component enters the model through $\pi f(\cdot \mid \sigma_i^2)$, the impact of approximating the true conditional density $f(\cdot \mid \sigma_i^2)$ by a common working density $\tilde{f}$ is limited when $\pi$ is small. We therefore expect the effect of moderate departures from the independence assumption on FDR control to be relatively limited when the non-null proportion is small. 

\begin{figure}[!htb]
    \centering
    \includegraphics[width=0.7\linewidth]{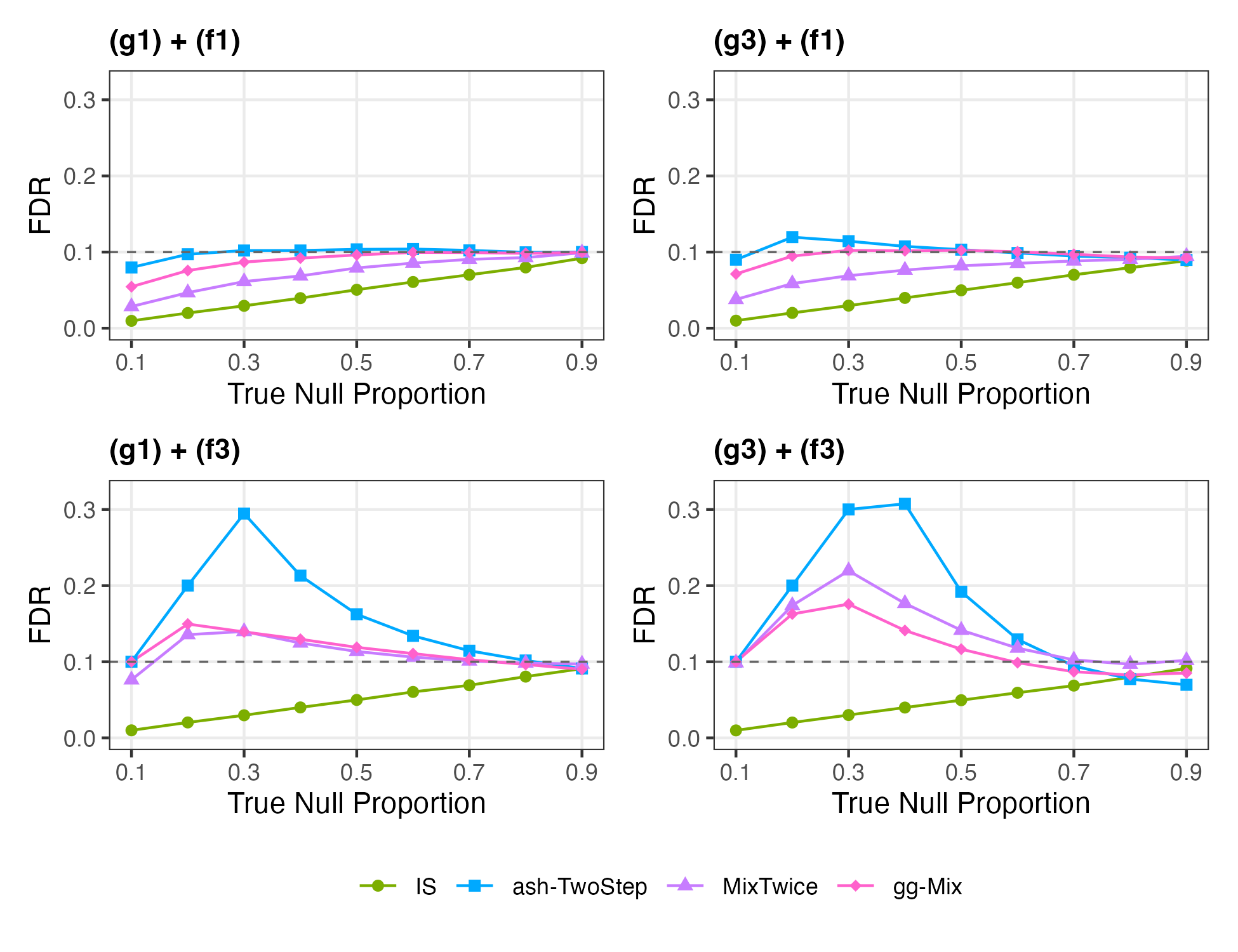}
    \caption{FDR under dependence between the effect size $\mu_i$ and variance $\sigma_i^2$.}
    \label{fig:F1}
\end{figure}

\begin{figure}[!htb]
    \centering
    \includegraphics[width=0.7\linewidth]{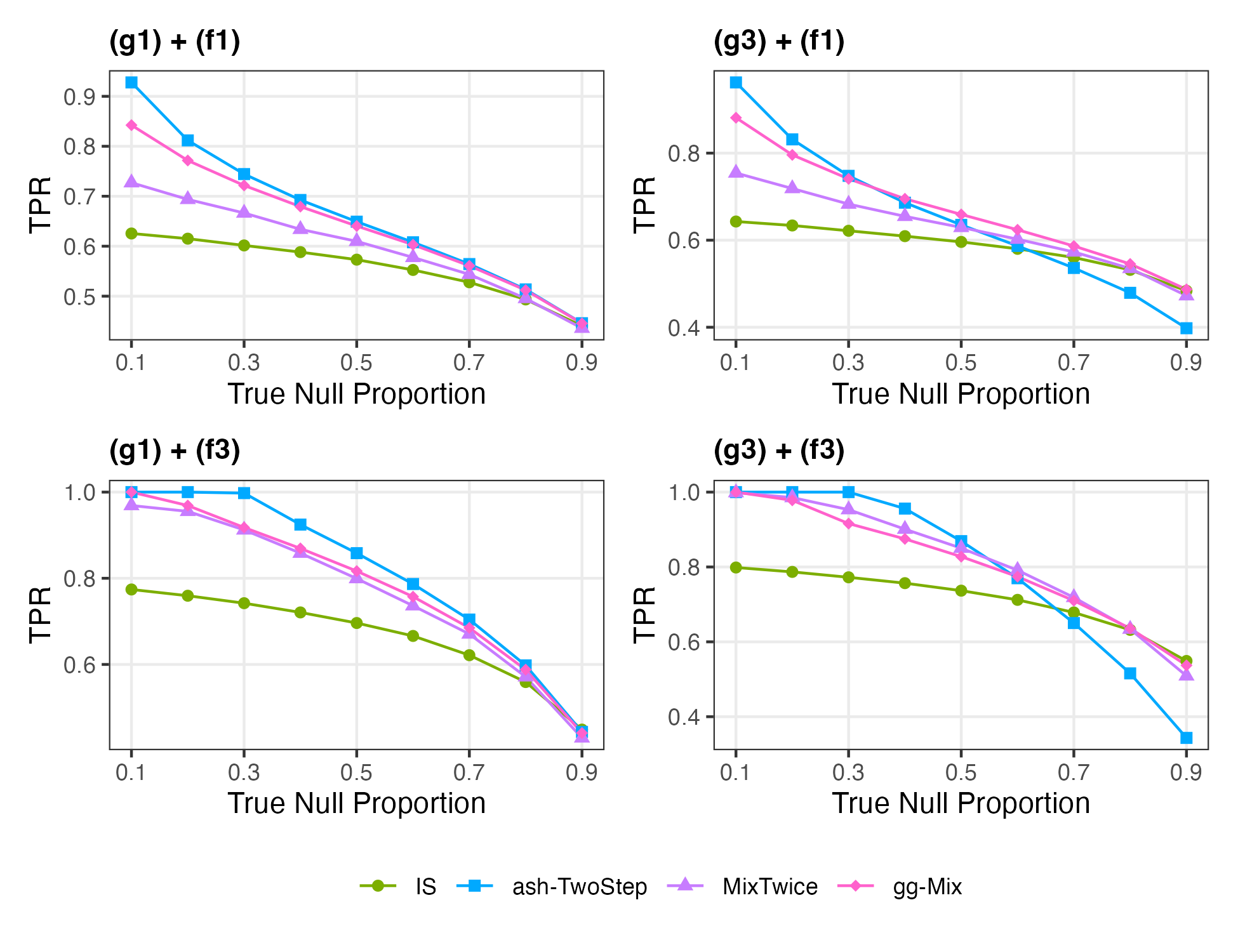}
    \caption{TPR under dependence between the effect size \(\mu_i\) and variance \(\sigma_i^2\).}
    \label{fig:F2}
\end{figure}

To examine empirically the effect of violating the independence assumption between $\mu_i$ and $\sigma_i^2$, we conduct an additional numerical study. We consider a setting in which 
\[
\mu_i/\sigma_i \sim (1-\pi)\delta_0 + \pi f
\]
in place of $\mu_i$, so that the magnitude of the effect size scales with the standard deviation $\sigma_i$. We consider $(\mathrm{g1})$ and $(\mathrm{g3})$ for $G$ and $(\mathrm{f1})$ and $(\mathrm{f3})$ for $f$, while varying $\pi_0$ from $0.1$ to $0.9$. 
The remaining settings are $m=5{,}000$, $\nu=10$, and a target FDR level of $0.1$. The FDR and TPR are each averaged over 200 replications. The results are summarized in Figures~\ref{fig:F1}--\ref{fig:F2}. Note that, in all panels, the x-axis represents the null proportion $\pi_0$, so a smaller value of $\pi_0$ corresponds to a larger non-null proportion $\pi$. Since our primary interest is the impact of dependence on the proposed procedure, we focus our interpretation on gg-Mix. 

Under $(\mathrm{f1})$, the FDR remains controlled at the nominal level of $0.1$ across all values of the null proportion. Under $(\mathrm{f3})$, in contrast, the misspecification appears to act in an anti-conservative direction, leading to FDR inflation as the non-null proportion increases. Interestingly, the FDR decreases again as the null proportion approaches $0.1$. Although this may appear counterintuitive, it is a natural consequence of having very few null hypotheses. When $\pi_0=0.1$, even rejecting all hypotheses yields an FDP of at most $0.1$.

In terms of TPR, for null proportions of $0.5$ or greater, a range that is more relevant to typical large-scale multiple testing applications, gg-Mix achieves comparable or higher power than the other methods that control the FDR at the nominal level.

\subsection{Violation of Independence Across $\{(X_i,S_i^2)\}_{i=1}^m$}

Dependence across observations is also common in real-data applications. To examine how gg-Mix responds to dependence across observations, we conduct an additional numerical study. We maintain the other assumptions of the hierarchical model in \eqref{eqn:assum_prior_1}--\eqref{eqn:like_2} while introducing dependence only across the signal indicators. For this purpose, we use a hidden Markov model (HMM), a standard framework for representing local dependence and clustering of signals in large-scale multiple testing. 

\begin{figure}[!htb]
    \centering
    \includegraphics[width=\linewidth]{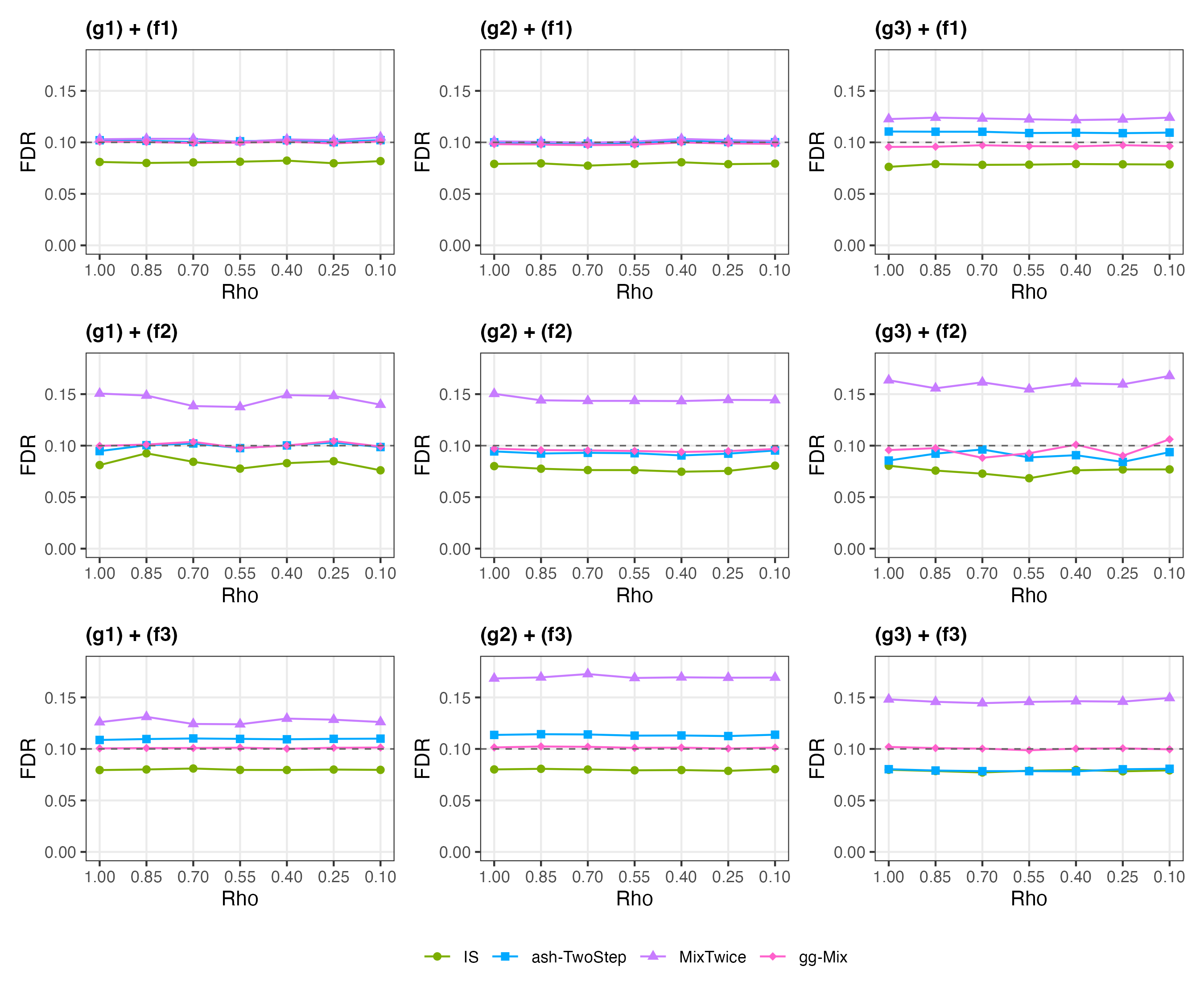}
    \caption{FDR under locally dependent signal configurations induced by the hidden Markov model.}
    \label{fig:F3}
\end{figure}

\begin{figure}[!htb]
    \centering
    \includegraphics[width=\linewidth]{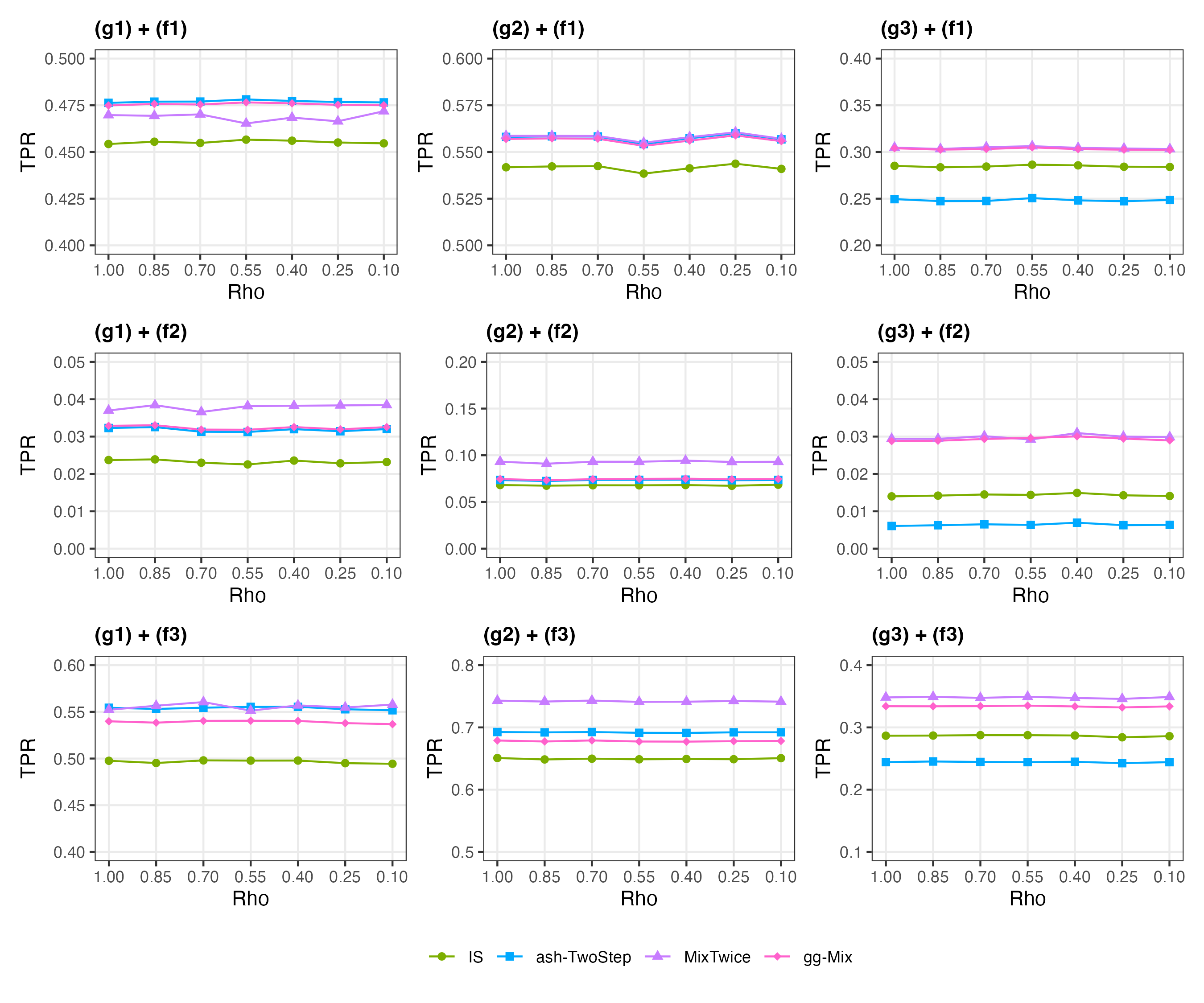}
    \caption{TPR under locally dependent signal configurations induced by the hidden Markov model.}
    \label{fig:F4}
\end{figure}

Specifically, we assume that the indicators $\theta_i = \mathbb{I}\{\mu_i \neq 0\}$ follow a two-state Markov chain, so that the pairs $(X_i, S_i^2)$ are no longer marginally independent across $i$. The transition probabilities are parameterized as
\[
    P(\theta_i = 1 \mid \theta_{i-1} = 0) = \rho(1-\pi_0), \qquad
    P(\theta_i = 0 \mid \theta_{i-1} = 1) = \rho\pi_0.
\]
Under this parameterization, the stationary distribution is $(\pi_0,1-\pi_0)$ for every $\rho$, while the correlation between adjacent indicators is $1-\rho$. Thus, $\rho=1$ corresponds to the independent case, whereas smaller values of $\rho$ induce stronger clustering of the signals. We vary $\rho$ from $1$ to $0.1$, corresponding to adjacent correlations ranging from $0$ to $0.9$. The remaining setting consists of $m=5{,}000$ hypotheses with $\nu=10$, $\pi_0=0.8$, and a target FDR level of $0.1$. The results are summarized in Figures~\ref{fig:F3}--\ref{fig:F4}.

Across all values of $\rho$ considered, the FDR of gg-Mix remains controlled at the target level, while the TPR remains essentially unchanged. This pattern holds across all combinations of $G$ and $f$ considered. These results suggest that, under the dependence structure considered here, gg-Mix is robust to increasing local dependence, even when the correlation between adjacent signal indicators is as high as $0.9$.

We emphasize, however, that this numerical study considers only one specific form of dependence, namely local clustering of the signal positions induced by the HMM. The results therefore should not be interpreted as establishing FDR control under general dependence structures. Investigating the behavior of gg-Mix under broader forms of dependence, and developing methods that explicitly accommodate such dependence, remain important directions for future research.

\section{Semi-Synthetic Analysis}\label{appendix_g}
In this section, we conduct an additional semi-synthetic analysis to evaluate FDR control and power in a setting where the true null and non-null status of each hypothesis is known, while preserving important characteristics of real gene-expression data. For this analysis, we use the multiple myeloma microarray data of \citet{tian2003role}. The dataset contains gene-expression measurements from 173 multiple myeloma patients, including 36 patients without focal bone lesions and 137 patients with focal lesions. Our analysis uses expression measurements for 12{,}625 genes.

We adopt the semi-synthetic framework proposed by \citet{lu2019empirical}. To construct each semi-synthetic dataset, we first generate two artificial groups from the same population. Specifically, we use the 137 patients with focal lesions as a common pool, sample 20 patients from this pool, and randomly assign 10 patients to each of the two groups. Because both groups are constructed from the same patient population, there is no systematic difference in gene expression between the two groups before adding signals.

To control the degree of cross-gene dependence retained from the original data, the patient assignment is performed in blocks of genes of size $b\in\{1,10,100\}$. All genes within the same block share the same patient assignment, whereas assignments are generated independently across blocks. Thus, $b=1$ largely disrupts the original cross-gene dependence, while larger
values of $b$ retain the dependence structure within increasingly larger blocks of genes.

Importantly, because the expression measurements are sampled directly from the original data, this construction preserves gene-specific heteroscedasticity. It also retains the cross-gene dependence within each block, with the block size $b$ controlling the extent to which this dependence is preserved.

\begin{figure}[!htb]
    \centering
    \includegraphics[width=\linewidth]{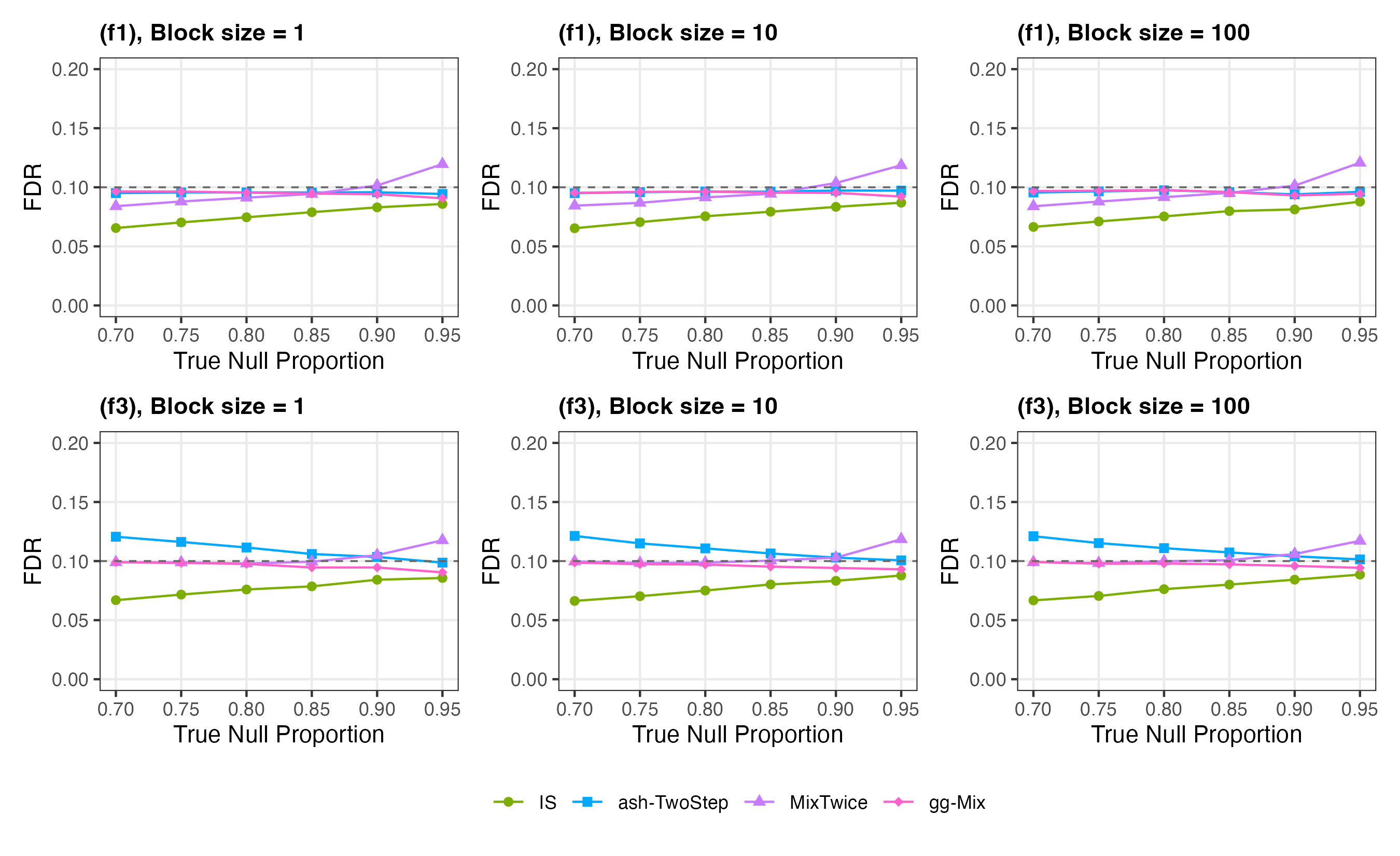}
    \caption{FDR in the semi-synthetic multiple myeloma data analysis under varying null proportions and degrees of cross-gene dependence.}
    \label{fig:G1}
\end{figure}

\begin{figure}[!htb]
    \centering
    \includegraphics[width=\linewidth]{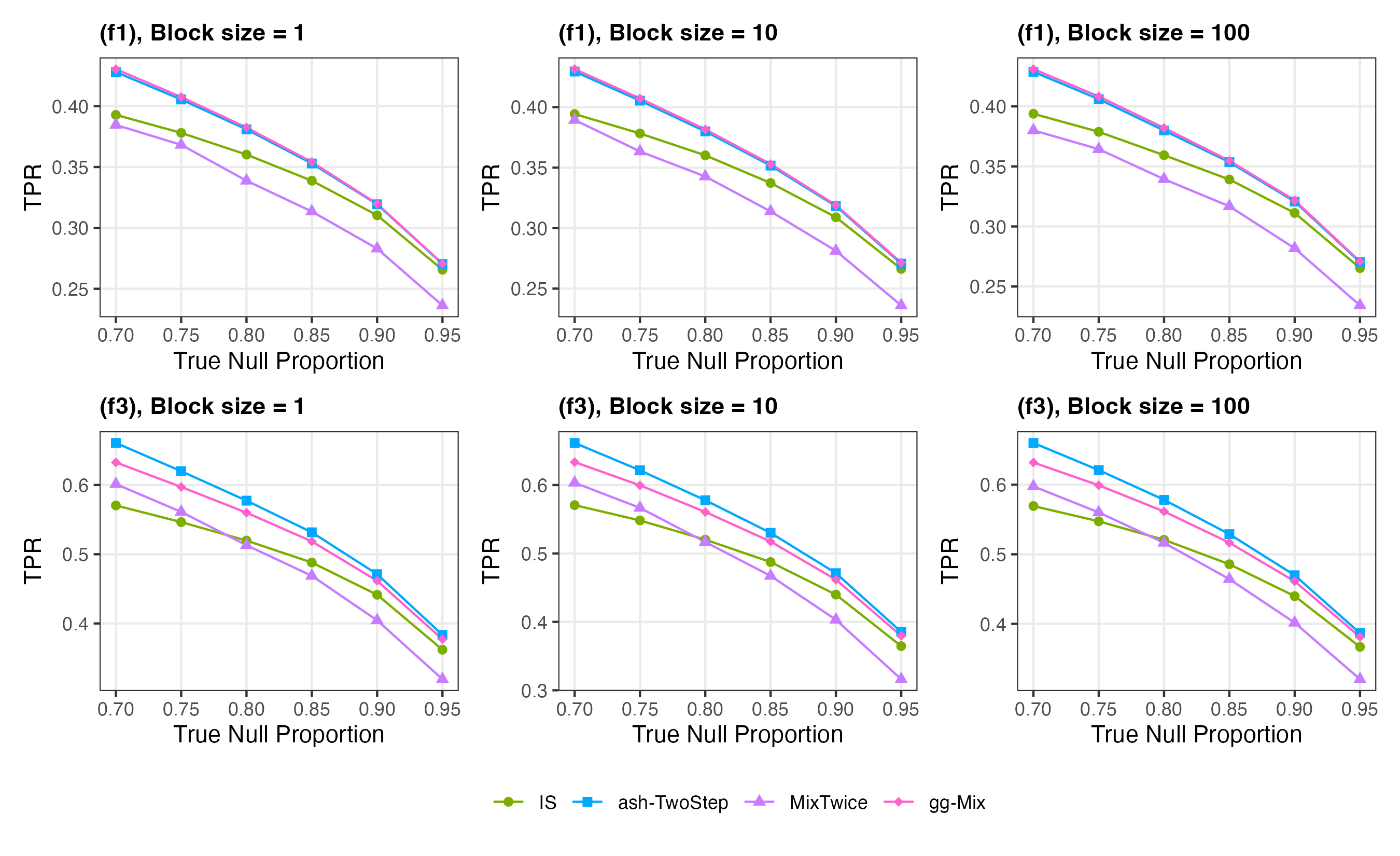}
    \caption{TPR in the semi-synthetic multiple myeloma data analysis under varying null proportions and degrees of cross-gene dependence.}
    \label{fig:G2}
\end{figure}

We then randomly designate a proportion $1-\pi_0$ of the genes as non-null and add known signals to one of the two artificial groups. The remaining genes are left unchanged and treated as null hypotheses. In this way, the null and non-null states are known by construction, allowing us to directly evaluate the FDR and TPR. We consider null proportions $\pi_0\in\{0.70,0.75,0.80,0.85,0.90,0.95\}$. For the remaining settings, we consider (f1) and (f3) for the non-null effects and fix the signal-to-noise ratio at 3. The target FDR level is set to $0.1$. The results are summarized in Figures~\ref{fig:G1}--\ref{fig:G2}. We observe the following patterns, which are broadly consistent with the results of our numerical studies in the main manuscript:

\begin{enumerate}
    \item[(a)] The FDR and TPR curves are nearly identical across the three block sizes, indicating that the performance of all methods is largely unaffected by the degree of cross-gene dependence retained in this semi-synthetic setting.

    \item[(b)] gg-Mix and IS control the FDR at the target level of 0.1 across all settings considered. However, gg-Mix consistently achieves substantially higher TPR than IS.
    
    \item[(c)] ash-TwoStep controls the FDR when the non-null effects are generated from (f1), where its unimodality assumption is satisfied. However, when the non-null effects are generated from the bimodal distribution (f3), ash-TwoStep fails to control the FDR in several settings. Although ash-TwoStep attains the highest TPR under (f3), this higher power is accompanied by FDR inflation and therefore does not represent a valid power advantage over gg-Mix. 
    
    \item[(d)] MixTwice generally controls the FDR but exhibits FDR inflation at large null proportions, particularly when $\pi_0=0.95$. Moreover, MixTwice consistently shows lower TPR than gg-Mix across the settings considered.
\end{enumerate}

\end{document}